\documentclass[11pt,a4paper]{article}
\pdfoutput=1 
\usepackage{jheppub} 

\usepackage{float}
\usepackage{geometry} 
\usepackage{setspace} 
\usepackage{lipsum}    

\usepackage{mathtools} 
\usepackage{xcolor}   
\usepackage{mathrsfs}  
\usepackage{leftidx}  
\usepackage{arydshln}  
\usepackage{bm} 
\usepackage{shuffle}   
\usepackage{slashed} 

\usepackage{tikz}

\numberwithin{equation}{section} 

\renewcommand{\eqref}[1]{(\ref{#1})}

\newcommand{\bg}[0]{\overline{g}} 
\newcommand{\softS}[1]{S^{(0)}_{#1}}
\newcommand{\halfsoftS}[1]{\frac{\delta_{#1}}{s_{#1}}} 
\newcommand{\subsoftS}[1]{S^{(1)}_{#1}}
\newcommand{\ba}[0]{\bm{\alpha}}
\newcommand{\bb}[0]{\bm{\beta}}
\newcommand{\bs}[0]{\bm{\sigma}}

\newcommand{\br}[0]{\bm{\rho}}
\newcommand{\bvr}[0]{\bm{\varrho}}

\newcommand{\bga}[0]{\bm{\gamma}} 
\newcommand{\bth}[0]{\bm{\theta}}
\newcommand{\kPreN}[0]{\mathcal{N}_{\mathbb{I}}}
\newcommand{\leftX}[1]{X^{L}_{#1}} 
\newcommand{\rightX}[1]{X^{R}_{#1}} 
\newcommand{\leftOrderX}[2]{X^{L(#2)}_{#1}} 
\newcommand{\rightOrderX}[2]{X^{R(#2)}_{#1}} 
\newcommand{\leftRhoX}[1]{\leftOrderX{#1}{\bm{\rho}}} 
\newcommand{\rightRhoX}[1]{\rightOrderX{#1}{\bm{\rho}}} 

\newcommand{\yms}[0]{A_{\text{YMS}}}
\newcommand{\bas}[0]{A_{\text{BS}}}
\newcommand{\llangle}[0]{\left\langle}
\newcommand{\rrangle}[0]{\right\rangle}

\allowdisplaybreaks

\title{Note on the Hopf-algebra-based formula of Yang-Mills-Scalar amplitudes}

\author[a]{Jiexi Liu} 
\author[a,b]{Yi-Jian Du}

\affiliation[a]{Department of Physics, School of Physics and Technology,
Wuhan University, \\
No.299 Bayi Road, Wuhan 430072, China}
\affiliation[b]{Hubei Key Laboratory of Nuclear Solid Physics, School of Physics and Technology, Wuhan University,\\
No.299 Bayi Road, Wuhan 430072, China}

\emailAdd{jaywhu@whu.edu.cn} 
\emailAdd{yijian.du@whu.edu.cn}

\date{\today}

\abstract{
In this note, we study the Hopf-algebra-based (HAB) formula of Yang-Mills-Scalar (YMS) amplitudes, which expands a YMS amplitude with massive scalars as a combination of propagator matrices that mix massless scalars corresponding to gluons with the original massive scalars. We propose a recursive formula which conveniently expresses the HAB formula. In this formula, gluons are converted into massless scalars. Thus it expresses a YMS amplitude with massive scalars by amplitudes with fewer gluons, massive scalars and massless scalars. We verify this formula by using soft behavior approach. In the massless limit, the HAB formula turns into expressions for YMS amplitudes with massless scalars which was earlier shown to satisfy an alternative recursive expansion formula. In this note, we show the equivalence of these two distinct approaches through explicit calculation on amplitudes with one and two gluons.
}

\begin{document}
\maketitle \flushbottom

\newpage

\section{Introduction}

The study of scattering amplitudes in Yang-Mills-scalar (YMS) theory plays a crucial role in understanding the perturbative relation between Yang-Mills (YM) theory and general relativity (GR). The reason is demonstrated as follows. Once YMS amplitudes are expanded in terms of bi-adjoint scalar (BS) \cite{Cachazo:2013iea} amplitudes, one can immediately write down an expression of Yang-Mills amplitude, also written in terms of BS ones \cite{expEYM}. The coefficients in this formula, have been used to generate all Bern-Carrasco-Johansson (BCJ) \cite{BCJBasis,Bern:2010ue} numerators \cite{Du:2017kpo}, which are the key objects in the double-copy relation between YM and GR.

A concrete realization of the expansion of massless-scalar YMS amplitudes into BS ones is the recursive expansion formula \cite{expEYM}\footnote{More on this recursive formula can be found in \cite{Chiodaroli:2017ngp,Teng:2017tbo,Du:2017gnh,Hou:2018bwm,Du:2019vzf}} which is based on the manifesting of gauge invariance. According to this approach, any tree level single-trace YMS amplitude (with massless scalars) are expressed into a combination of YMS amplitudes with fewer gluons. This formula, when applied iteratively, finally results in a BS expansion of YMS amplitudes. It have been show that the recursive expansion relation are effective for the construction of tree-level Bern-Carrasco-Johansson (BCJ) \cite{BCJBasis,Bern:2010ue} numerators. They are also generalized to one-loop integrands by the forward limit strategy \cite{He:2016mzd,He:2017spx,Geyer:2015jch,Geyer:2017ela,Edison:2020uzf,Porkert:2022efy}. An interesting perspective is that the recursive expansion formula can be determined by the universal soft behaviors, see \cite{Zhou:2022orv,Zhou:2023iae,Hu:2023koy,Du:2024xva}.

Hopf-algebra-based (HAB) formula \cite{Chen:2022nei} is an alternertive approach to the YMS amplitude (but with general massive scalars). This formula originated from the study of heavy-mass effective field theory \cite{Brandhuber:2021kpo,Brandhuber:2021bsf,HopfHEFT}. By the help of HAB formula, YMS amplitudes with {\it massive scalars}\footnote{In this note, all massive scalars are supposed to have the same mass.} are expressed in terms of scalar propagator matrices combining massless scalars corresponding to gluons and the original massive scalars. The expansion coefficients are generated by a systematic rule based on the framework of Hopf algebra. By the help of double-copy statement, the HAB approach has been sucessfully applied in the study of massive particles scattering against gravitons \cite{HopfHEFT, HopfNote, HopfYMS, Bjerrum-Bohr:2024fbt, Chen:2023ekh, Chen:2024gkj} and have been used to calculate classical gravitational process \cite{Bjerrum-Bohr:2023jau, Brandhuber:2023hhy, Bjerrum-Bohr:2023iey, Brandhuber:2023hhl, Chen:2024mmm, Brandhuber:2024qdn}.

In this note, we study the HAB formula of YMS amplitudes. We arrange the HAB formula for YMS amplitudes with massive scalars into a convenient recursive relation, through which, a YMS amplitude is written as a combination of amplitudes\footnote{These `ampltiudes' are intemediate objects. We call them amplitudes in the sense that (i) the massless limit gives the ususal YMS amplitudes with massless scalars, (ii) they are gauge invariant. } mixing massive scalars, massless scalars and fewer gluons. The HAB formula can be regarded as an iterative result of this recursive formula. We verify this formula by the soft limit approach. We further prove that the HAB formula in the massless limit is equivalent to the other approach  \cite{expEYM}, by explicit calculations on amplitudes with one and two gluons.

The structure of the paper is organized as follows. In section \ref{sec:review}, we briefly review the recursive expansion formula proposed in \cite{expEYM}, the HAB formula \cite{Chen:2022nei} as well as the construction of amplitudes bases on universal soft behaviors \cite{Zhou:2022orv,Zhou:2023iae,Hu:2023koy,Du:2024xva}. The soft behavior of amplitudes is also introduced specifically for YMS amplitudes with massive scalars. In section \ref{sec:exp-from-hopf}, we construct a recursive version of the HAB formula. A verification of this recursive formula is provided in section \ref{sec:exp-from-soft}, based on the soft behavior approach.  In section \ref{sec:massless-expansion}, we instead derive a recursive expansion of YMS amplitudes involving only massless scalars, starting from the known formula \cite{expEYM}. Through explicit calculation, we demonstrate that this expansion matches the massless limit of the HAB formula for one- and two-gluon cases. Finally, we summarize this work in section \ref{sec:conclusion}. Helpful relations are included in the appedix.

\section{A review of recursive expansion, HAB formula and soft behaviors} \label{sec:review}
This section provides a review of two distinct approaches to the YMS amplitudes, the recursive expansion \cite{expEYM} formula and the Hopf-algebra-based approach \cite{HopfHEFT, HopfNote, HopfYMS, Bjerrum-Bohr:2024fbt, Chen:2023ekh, Chen:2024gkj}. Furthermore, we review the soft behavior construction of amplitudes, which plays as a role in this paper.

\subsection{The recursive expansion formula of YMS} \label{sec:review-exp}
As proposed in \cite{expEYM}, {\it massless} EYM amplitudes satisfy the recursive expansion formula, which also holds for YMS amplitudes. Concretely, for a tree-level doubly color-ordered massless YMS amplitude \cite{expEYM} involving $n$ scalars ($1,2, \ldots, n$) and $m$ gluons denoted by $\{g_i\}$, and a given right permutation $\bs$ of all scalars and gluons, we have
\begin{equation} \label{eq:exp-poly}
    \begin{aligned}
        \yms(1, \ldots, n \|  \{g_i\}  \, | \, \bs) = & \, \left(\epsilon_f \cdot \leftX{f}\right) \yms(1, \{2, \ldots, n-1\} \shuffle g_f, n  \|  \{g_i\} \! \setminus \! g_f  \, | \, \bs) \\
        & + \sum_{\ba}\left(\epsilon_f \cdot F_{\ba^{\text{T}}} \cdot \leftX{\ba}\right) \yms(1, \{2, \ldots, n-1\} \shuffle \{\ba, g_f\}, n  \|  \{g_i\} \! \setminus \! (g_f \cup \ba )  \, | \, \bs),
    \end{aligned}
\end{equation}
in which, $g_f$ is an arbitrarily chosen gluon ({\it fiducial gluon}), and  all non-empty ordered subsets $\ba = (\alpha_1, \ldots, \alpha_a)$ of gluons $\{g_i\} \! \setminus \! g_f$ on the second line has been summed over. The shuffle symbol $\shuffle$ denotes the sum over all possible permutations of scalars $\{2, \ldots, n-1\}$ and gluons $\{\ba, g_f\}$ preserving the relative ordering in each ordered set. For example, $\{2, 3\} \shuffle \{g_1, g_f\}$ implies that we sum over the following permutations
\begin{equation}
    \{2, 3\} \shuffle \{g_1, g_f\} : \{g_1, g_f, 2,3\},\{g_1, 2, g_f ,3\},\{g_1, 2, 3, g_f\},\{2, g_1, g_f, 3\}, \{2,g_1, 3, g_f\}, \{2,3,g_1,g_f\}.
\end{equation}
The  $\big(\leftX{f}\big)^{\mu}$ is the sum of the momenta of scalars on the left-hand side of the fiducial gluon $g_f$ in a given permutation of scalars and gluon $g_f$. Similarly, $\big(\leftX{\ba}\big)^{\mu}$ is the total momentum of scalars on the left-hand side of gluon $\alpha_1$ (the \textbf{leftmost} element in $\ba$) in a given ordering of scalars and gluons $\{ \ba, g_f \}$. The tensor $F^{\mu\nu}_{\ba^{\text{T}}}$ denotes the contraction of strength tensors $F^{\mu \nu} = k^\mu \epsilon^\nu - \epsilon^\mu k^\nu$ of gluons in $\ba$ and takes the form as 
\begin{equation} \label{}
    F_{\ba^{\text{T}}}^{\mu \nu} = (F_{\alpha_a} \cdot \ldots \cdot F_{\alpha_1})^{\mu \nu},
\end{equation}
where the ordering $\ba^{\text{T}} = (\alpha_a, \ldots, \alpha_1)$ is the inverse of $\ba = (\alpha_1, \ldots, \alpha_a)$.

Applying (\ref{eq:exp-poly}) iteratively, a doubly color-ordered YMS amplitude can be expanded into pure color-ordered BS ones
\begin{equation} \label{eq:exp-BS-KK}
    \yms(1, \ldots, n \|  \{g_i\}  \, | \, \bs) = \sum_{\bm{\br}} C(\br) A_{\text{BS}}(1, \br, n  \, | \, \bs),
\end{equation}
with $1$ and $n$ fixed as the two ends. Here we have summed over all possible permutations of the scalars and gluons preserving the scalar ordering $\{2, \ldots, n-1\}$. Each coefficient $C(\br)$ arises from the corresponding class of expansion routes, which leads to the specific BS amplitude $A_{\text{BS}}(1, \br, n  \, | \, \bs)$. For example, YMS amplitude with two scalars and two gluons $\yms(1, 2  \|  \{g_1, g_2\}  \, | \, \bs)$ can be expanded with the recursive expansion formula (\ref{eq:exp-poly}) applied repeatedly
\begin{align}
    \yms(1, 2  \|  \{g_1, g_2\}  \, | \, \bs) & = (\epsilon_1 \cdot p_1) \yms(1, g_1, 2  \|  g_2  \, | \, \bs) + (\epsilon_1 \cdot F_2 \cdot p_1) \bas(1, g_2, g_1, 2  \, | \, \bs), \\
    \yms(1, g_1, 2  \|  g_2  \, | \, \bs) & = \left[ \epsilon_2 \cdot (p_1 + k_1) \right]\bas(1, g_1, g_2, 2  \, | \, \bs) + (\epsilon_2 \cdot p_1) \bas(1, g_2, g_1, 2  \, | \, \bs),
\end{align}
where we denote the momentum of scalar $i$ by $p_i$ and that of gluon $j$ by $k_j$. The second line shows a further expansion of YMS amplitude $\yms(1, g_1, 2  \|  g_2  \, | \, \bs)$, which is on the RHS of the first expansion. Two expansions are combined to derive the expansion to pure BS amplitudes. According to the specific BS amplitude $\bas(1, \br, 2  \, | \, \bs)$, we can rearrange the expansion coefficients from the corresponding expansion routes into $C(\br)$. The coefficient $C(g_1,g_2)$ arises from the two-step expansion route $(1,2) \to (1, g_1 ,2) \to (1, g_1, g_2, 2)$, while $C(g_2,g_1)$ collects the contributions from  $(1, 2) \to (1, g_2, g_1, 2)$ and the two-step expansion route $(1,2) \to (1, g_1 ,2) \to (1, g_2, g_1, 2)$. Concretely, we have
\begin{equation}
    \begin{aligned}
        C(g_1, g_2) = & \, (\epsilon_1 \cdot p_1) \left[\epsilon_2 \cdot (p_1 + k_1) \right], \\
        C(g_2, g_1) = & \, (\epsilon_1 \cdot F_2 \cdot p_1) + (\epsilon_1 \cdot p_1) (\epsilon_2 \cdot p_1),
    \end{aligned}
\end{equation}
and the explicit expression of the expansion of YMS amplitudes to pure BS amplitudes as
\begin{equation}
    \yms(1, 2  \|  \{g_1, g_2\}  \, | \, \bs) = C(g_1, g_2) \bas(1, g_1, g_2, 2  \, | \, \bs) + C(g_2, g_1) \bas(1, g_2, g_1, 2  \, | \, \bs).
\end{equation}
As shown by \cite{expEYM}, this formula also holds for amplitude with a pair of massive scalars.

\subsection{Hopf-algebra-based formula} \label{sec:reviewHopf}
The Hopf algebra framework \cite{HopfHEFT, HopfNote, HopfYMS, Bjerrum-Bohr:2024fbt, Chen:2023ekh, Chen:2024gkj} provides a systematic way to express the scattering amplitudes involving gluons and massive scalars or fermions. In this framework, the color ordering \footnote{In \cite{HopfYMS}, mentioned as flavor orderings. To agree with the convention of many works on massless YMS amplitudes, we adopt `color ordering' instead. We hope this will not bring any misuderstanding.} is kept in the theory-independent Hopf algebra structure, while the kinematic data of amplitudes is collected in the theory-dependent linear map. In this paper, we focus on the YMS amplitudes with massive scalars (where all scalar particles have the same mass) and massless gluons. 

As proposed in \cite{HopfYMS}, single-colored YMS amplitudes can be represented as
\begin{equation} \label{eq:hopf-BCJ}
    \yms(\bm{\varsigma}, n) = \sum_{\Gamma \in R_{\bm{\varsigma}}} \frac{\llangle \widehat{\mathcal{N}}(\Gamma) \rrangle}{d_\Gamma}.
\end{equation}
Here $\bm{\varsigma}$ denotes the ordering of gluons and scalars with scalar $n$ fixed due to the cyclic symmetry. We sum over all possible tri-graphs $\Gamma$ following the ordering $\bm{\varsigma}$. Each tri-graph $\Gamma$ corresponds to a unique propagator in the denominator, and implies the nest commutator structure in the numerator. The following examples illustrate the tri-graphs along with their corresponding numerators and denominators.
\begin{equation} \label{eq:tri-graph-example1}
    \Gamma: \vcenter{\hbox{
        \begin{tikzpicture}[
        thick,
        dot/.style={circle, fill=black, inner sep=2.2pt} 
        ]
            \node[dot] (v12) at (0, 0) {};
            \node[dot] (v1g1) at (-0.5, 0.7) {};

            \draw (v12) -- (v1g1);

            \draw (v12) -- (0, -1);      
            \draw (v12) -- (1, 1.4);     

            \draw (v1g1) -- (-1, 1.4);      
            \draw (v1g1) -- (-0.1, 1.4);    

            \node[below] at (0, -1.1) {$3$};

            \node at (-1.05, 1.6) {$1$};
            \node at (-0.15, 1.6) {$g_1$};
            \node at (1.05, 1.6) {$2$};  
        \end{tikzpicture}
    }} \rightarrow \quad \frac{\llangle \widehat{\mathcal{N}}(\Gamma) \rrangle}{d_\Gamma} = \frac{[[K_{1},K_{g_1}],K_{2}]}{(s_{1 g_1} - m_s^2)(s_{12 g_1} - m_s^2)},
\end{equation}

\begin{equation} \label{eq:tri-graph-example2}
    \Gamma: \vcenter{\hbox{
        \begin{tikzpicture}[
        thick,
        dot/.style={circle, fill=black, inner sep=2.2pt} 
        ]
            \node[dot] (v12) at (0, 0) {};
            \node[dot] (v1g1) at (-0.34, 0.46) {};
            \node[dot] (vg1g2) at (-0, 0.92) {};

            \draw (v12) -- (v1g1);
            \draw (vg1g2) -- (v1g1);

            \draw (v12) -- (0, -1);      
            \draw (v12) -- (1, 1.4);     

            \draw (v1g1) -- (-1, 1.4);      

            \draw (vg1g2) -- (-0.34, 1.4);
            \draw (vg1g2) -- (0.34, 1.4);

            \node[below] at (0, -1.1) {$3$};

            \node at (-1.05, 1.6) {$1$};
            \node at (-0.34, 1.6) {$g_1$};
            \node at (0.34, 1.6) {$g_2$};
            \node at (1.05, 1.6) {$2$};
        \end{tikzpicture}
    }} \rightarrow \quad \frac{\llangle \widehat{\mathcal{N}}(\Gamma) \rrangle}{d_\Gamma} = \frac{[[K_{1},[K_{g_1},K_{g_2}]],K_{2}]}{s_{g_1 g_2}(s_{1 g_1 g_2} - m_s^2)(s_{12 g_1 g_2} - m_s^2)}.
\end{equation}
Here, the Mandelstam variables $s_{i \ldots g_j \ldots }\equiv (p_i + \ldots + k_j + \ldots)^2$ (where $p_i, k_j, \ldots$ denote the momenta of particles) are used to simplify expressions. The commutator of the single-particle generators is defined as $[K_i,K_j] = K_i \star K_j - K_j \star K_i$, while for the details of the fusion product $\star$ one can refer to \cite{HopfHEFT, HopfNote, HopfYMS}. Since each numerator is defined as the properly nested commutator determined by the corresponding tri-graph, these numerators satisfy the Jacobi identity and thus qualify as BCJ numerators \cite{HopfYMS}.

The above singly color-ordered YMS amplitude (\ref{eq:hopf-BCJ}) are connected to doubly color-ordered amplitudes in (\ref{eq:exp-poly}) and (\ref{eq:exp-BS-KK}) via the color decomposition
\begin{equation}
    \yms(\bm{\varsigma}, n) = \sum_{\bm{\eta} \in S_{n-1}} \yms(\bm{\eta}, n \|  \{g_i\}  \, | \, \bm{\varsigma}, n) \operatorname{tr}(t^{\bm{\eta}} t^n).
\end{equation}
On the LHS, $\bm{\varsigma}$ is a permutation of all elements in $\{1, \ldots, n-1\} \cup \{g_i\}$. On the RHS, we use $\operatorname{tr}(t^{\bm{\eta}} t^n)$ as shorthand for $\operatorname{tr}(t^{\eta_1} \ldots t^{\eta_{n-1}} t^n)$, where each $\bm{\eta}=\{\eta_1, \ldots, \eta_{n-1}\}$ is a permutation of scalars $\{1, \ldots, n-1\}$.

Considering that the propagator structure (\ref{eq:tri-graph-example1}, \ref{eq:tri-graph-example2}) of tri-graphs can be captured by the propagator matrix, an equivalent representation of YMS amplitudes was derived \cite{HopfYMS} from the expression (\ref{eq:hopf-BCJ}) introduced above. This is essentially the expansion of doubly color-ordered YMS amplitudes into the propagator matrix with kinematic coefficients 
\begin{equation} \label{eq:hopf-result}
    \yms(1, \ldots, n-1, n \|  \{g_i\}  \, | \, \bs) = \sum_{\br} \mathbf{m}(1, \br, n-1, n  \, | \, \bs) \, \kPreN\left(1, \br, n-1, n\right),
\end{equation}
where $\br = (\rho_1, \ldots, \rho_{n+m-3})$ denotes the permutation of $(n-3)$ scalars $\{2, \ldots, n-2\}$ and $m$ gluons doubly color-ordered sum over all $\br$ preserving the scalar ordering $\br^s = \{2, \ldots, n-2\}$. The $\mathbf{m}(1, \br, n-1, n  \, | \, \bs)$ denotes the propagator matrix \cite{HopfYMS} involving the original massive scalars and massless scalars which are converted from gluons. For example, the propagtor matrix $\mathbf{m}(1, 2, g_1, g_2, 3  \, | \, 1, 2, g_1, g_2, 3)$, involving massive scalars $1, 2, 3$ and massless scalars corresponding to $g_1, g_2$, is given as 
\begin{equation}
    \begin{aligned}
        \mathbf{m}(1,2,g_1,g_2,3 \, \| \, 1,2,g_1,g_2,3)
        &= \frac{1}{(s_{12} - m_s^2)\,s_{g_1 g_2}}
        + \frac{1}{(s_{12} - m_s^2)\,(s_{1 2 g_1} - m_s^2)}
        + \frac{1}{(s_{2 g_1} - m_s^2)\,(s_{1 2 g_1} - m_s^2)} \\[4pt]
        &\quad+ \frac{1}{(s_{2 g_1} - m_s^2)\,(s_{2 g_1 g_2} - m_s^2)}
        + \frac{1}{s_{g_1 g_2}\,(s_{2 g_1 g_2} - m_s^2)} \,.
    \end{aligned}
\end{equation}

The kinematic coefficient $\kPreN(1, \br, n-1, n)$ in (\ref{eq:hopf-result}) is defined by applying the linear map $\llangle \cdot \rrangle_{\text{m}}$ to the fusion product $\llangle \mathrm{K}_{1} \star \mathrm{K}_{\br_1} \star \ldots \star \mathrm{K}_{\br_{n+m-3}} \star \mathrm{K}_{n-1} \rrangle_{\text{m}}$, which follows the order $\br$. For details, see \cite{HopfHEFT, HopfNote, HopfYMS}. The explicit expression of the kinematic coefficient $\kPreN(1, \br, n-1, n)$ is 
\begin{equation} \label{eq:kinematic-pre-numerator}
    \begin{aligned}
        \kPreN(1, \br, n-1, n) = & \, (-1)^{m} \sum_{r=1}^{m} (-1)^{r} \sum_{\ba^l \in  \mathbf{P}_{ \!\! \br^g}^{r}} \prod_{l=1}^{r} \frac{2 \leftRhoX{\ba^l} \cdot F_{\ba^l} \cdot \rightRhoX{\ba^l} }{P_l^2 - m_s^2}, \\
        = & \, \sum_{r=1}^{m} \sum_{\ba^l \in  \mathbf{P}_{ \!\! \br^g}^{r}} \prod_{l=1}^{r} - \frac{(-1)^{|\ba^l|} 2 \leftRhoX{\ba^l} \cdot F_{\ba^l} \cdot \rightRhoX{\ba^l} }{P_l^2 - m_s^2}.
    \end{aligned}
\end{equation}
Here, the second line is merely a rearrangement of the signature. The gluon ordering $\br^g$ inherits from $\br$. For example, the overall ordering $(1, \br, 3, 4) = (1, g_1, 2, g_2, 3, 4)$ implies gluon ordering $\br^g = (g_1, g_2)$ and preserves the scalar ordering $\br^s = (1, 2, 3)$ (including every scalar except $n$). We first sum, for fixed $r$, over all ordered partitions $\mathbf{P}_{ \!\! \br^g}^{r}$ that divide the gluon sequence $\br^g$ into $r$ subsets, each preserving $\br^g$, and then sum over the all possible $r$. For example, all ordered partitions of $\br^g = (g_1, g_2, g_3)$ are given as  
\begin{equation}
    \begin{aligned}
        & \mathbf{P}_{ \!\! \br^g}^{1}:\{(g_1,g_2,g_3)\}, \\
        & \mathbf{P}_{ \!\! \br^g}^{2}:\text{Perm}\{(g_1, g_2),(g_3)\}, \text{Perm}\{(g_1,g_3),(g_2)\}, \text{Perm}\{(g_2,g_3),(g_1)\}, \\
        & \mathbf{P}_{ \!\! \br^g}^{3}: \text{Perm}\{(g_1),(g_2),(g_3)\}.
    \end{aligned}
\end{equation}
Here $\text{Perm}$ denotes all possible permutations of the subsets. Notice that the ordering within each subset inherits from the overall ordering $\br^g = (g_1, g_2, g_3)$. The \textbf{denominators} in (\ref{eq:kinematic-pre-numerator}) are in the form of massive propagators with the scalar mass $m_s$. For specific $l$, $P_l$ is defined as the sum of momenta carried by all scalar particles (excluding $n$) and all gluons contained in the sets $\ba^1, \ldots, \ba^{l-1}$. In the \textbf{numerator}, the tensor $F_{\ba^l}^{\mu \nu}$ denotes the consecutive contraction of  strength tensors $(F_{\alpha^l_1} \cdot \ldots \cdot F_{\alpha^l_a})^{\mu \nu}$ preserving the gluon ordering $\ba^l = (\alpha^l_1, \ldots, \alpha^l_a)$. The $\leftRhoX{\ba^l}$ collects the momenta of the scalars in  $\br^s$ and gluons in $\ba^1, \ldots, \ba^{l-1}$ that lie to the left of gluon $\alpha^l_1$ with respect to $\br$, while $\rightRhoX{\ba^l}$ does the same for those particles on the right of gluon $\alpha^l_a$. For example, for the ordering $\br = (1, g_1, 2, g_2, 3)$, the following shows the terms corresponding to different partitions $\{\ba^l\}$
\begin{align}
    \{(g_1,g_2)\} & : \frac{2 p_1 \cdot F_1 \cdot F_2 \cdot p_3}{p_{123}^2 - m_s^2}, \\
    \{(g_1),(g_2)\} & : \frac{2 p_1 \cdot F_1 \cdot p_{23}}{p_{123}^2 - m_s^2} \frac{2 (p_{12}+k_1) \cdot F_2 \cdot p_3}{(p_{123} + k_1)^2 - m_s^2}.
\end{align}

The above expression (\ref{eq:hopf-result}) is suitable for YMS amplitudes with at least three scalars, since three scalars $1$, $n-1$, and $n$ are selected to be fixed. For YMS amplitudes with two scalars, the expansion of amplitudes \cite{HopfYMS} takes the form of
\begin{equation} \label{eq:hopf-result-2s}
    \yms(1, 2 \|  \{g_i\}  \, | \, \bs) = \sum_{\br} \mathbf{m}(1, \br, g_1, 2  \, | \, \bs) \kPreN\left(1, \br, g_1, 2\right).
\end{equation}
Here, the gluon $g_1$ is selected as the fiducial gluon, with its position held fixed. Concretely, the kinematic coefficient \cite{HopfYMS} is given as
\begin{equation} \label{eq:kinematic-pre-numerator-2s}
    \begin{aligned}
        \kPreN(1, \br, g_1, 2) = & \, (-1)^{m} \sum_{r=1}^{m} (-1)^{r} \sum_{\ba^l \in  \mathbf{P}_{\!\! \br^g \setminus g_1}^{r}} - \frac{2 p_1 \cdot F_1 \cdot F_{\ba^1} \cdot p_1}{P_1^2 - m^2_s} \prod_{l = 2}^{m} \frac{2 \leftRhoX{\ba^l} \cdot F_{\ba^l} \cdot \rightRhoX{\ba^l}}{P_l^2 - m^2_s},\\
        = & \sum_{r=1}^{m} \sum_{\ba^l \in  \mathbf{P}_{\!\! \br^g \setminus g_1}^{r}} \frac{ (-1)^{|\ba^1|+1} 2 p_1 \cdot F_1 \cdot F_{\ba^1} \cdot p_1}{P_1^2 - m^2_s} \prod_{l = 2}^{m} - \frac{ (-1)^{|\ba^l|} 2 \leftRhoX{\ba^l} \cdot F_{\ba^l} \cdot \rightRhoX{\ba^l}}{P_l^2 - m^2_s},
    \end{aligned}
\end{equation}
where we sum over all ordered partitions of gluons $\br^g \! \setminus \! g_1$ since $g_1$ is already considered as a scalar. $P_1 \equiv k_1 + p_1$ is defined as the momentum sum of scalar particle $1$ and gluon $g_1$. For $l>2$, $P_l$ is extended to include the momenta of all gluons in the sets $\ba_1$ through $\ba_{l-1}$.

In this paper, we present a recursive expansion formula of YMS amplitudes, based on the (Hopf-algebra-based) HAB formulas (\ref{eq:hopf-result}, \ref{eq:kinematic-pre-numerator-2s}).

\subsection{The soft behavior approach to YMS amplitudes} \label{sec:reviewSoft}
Scattering amplitudes satisfy the universal soft behavior, which states that an $n$-point amplitude is factorized into an $n-1$ point amplitude and a soft factor, when the momentum of a massless particle tends to zero. In this subsection, we review the soft scalar and soft gluon behaviors of BS and YMS amplitudes, which will be useful in the reconstruction of amplitudes in the coming sections \ref{sec:exp-from-soft}.

The leading soft behavior of BS amplitudes with  $p_i \to \tau p_i$ ($\tau\to 0$) is given by 
\begin{equation} \label{eq:softBS}
    \bas^{(0)_{i}}(1, \ldots, n  \, | \, \bs) = \softS{i} \bas(1, \ldots, i-1, \slashed{i}, i+1, \ldots, n  \, | \, \bs),
\end{equation}
where  $\tau$ has been absorbed into the leading soft factor for later convenience. The soft factor for scalar $s_i$ takes the form
\begin{equation} \label{eq:soft-factor-scalar}
    \softS{i} = \frac{1}{\tau} \left(\halfsoftS{(i-1) i} + \halfsoftS{i (i+1)}\right).
\end{equation}
Here the operator $\delta_{ij}$ is defined as
\begin{equation} \label{eq:delta}
    \delta_{ij} = \left\{ 
    \begin{array}{ll} 
        1 & \quad i \prec j \\ 
        -1 & \quad j \prec i \\
        0 & \quad i, j \,\, \text{\small not adjacent}
    \end{array} \right. ,
\end{equation}
which is determined by the relative order of scalars $i$ and $j$ in the given ordering $\bs$. The notation $i \prec j$ means that $i$ is on the left-hand side of $j$ in $\bs$. Similarly, the leading soft behavior of YMS amplitudes with a soft scalar $i$ is represented as
\begin{equation} \label{eq:softYMScalar}
    \yms^{(0)_{i}}(1, \ldots, n  \|  \{g_j\}  \, | \, \bs) = \softS{i} \yms(1, \ldots, i-1, \slashed{i}, i+1, \ldots, n  \|  \{g_j\}  \, | \, \bs),
\end{equation}
where the leading scalar soft factor is the same as that of BS amplitudes (\ref{eq:soft-factor-scalar}) due to the universality of the scalar soft factor.

Analogous to the soft behavior of scalars, the YMS amplitudes are factorized under the soft limit of gluons. Specifically, when gluon $g_i$ is the soft one, i.e., $k_i \to \tau k_i$, the leading and subleading behaviors of YMS amplitudes are presented by
\begin{equation} \label{eq:softYMSGluon}
    \yms(1, \ldots, n \|  \{g_j\}  \, | \, \bs) = \left[\softS{g_i} + \subsoftS{g_i} \right] \yms(1, \ldots, n \|  \{g_j\} \! \setminus \! g_i  \, | \, \bs \! \setminus \! g_i) + \mathcal{O}(\tau^1),
\end{equation}
The factors $\softS{g_i}$ and $\subsoftS{g_i}$ are leading and subleading gluon soft factors \cite{Cachazo:2014fwa,Casali:2014xpa}
\begin{equation} \label{eq:soft-factor-gluon}
    \softS{g_i} = \frac{1}{\tau} \sum_{a \neq g_i} \frac{\delta_{a g_i} (\epsilon_i \cdot p_a)}{s_{a g_i}}, \quad \subsoftS{g_i} = \sum_{a \neq g_i} \frac{\delta_{a g_i} (\epsilon_i \cdot J_a \cdot \mathsf{K}_{s_i})}{s_{a g_i}},
\end{equation}
where the operator $\delta_{a g_i}$ is defined similarly as $\delta_{ij}$ in (\ref{eq:delta}).
We sum over all external particles $a$ except for the soft gluon $g_i$ and introduce the angular momentum $J^{\mu \nu}_a$ for each particle $a$. It is appropriate to regard the subleading soft gluon factor as a differential operator acting on the kinematic variables of amplitudes excluding the gluon $g_i$. Explicitly, the soft factor \cite{Zhou:2022orv} acts in the form of 
\begin{align}
    \left(\subsoftS{g_i} k_a\right) \cdot V = & \, - \halfsoftS{a g_i} (k_a \cdot F_i \cdot V), \label{eq:soft-factor-on-k} \\
    \left(\subsoftS{g_i} \epsilon_a\right) \cdot V = & \, - \halfsoftS{a g_i} (\epsilon_a \cdot F_i \cdot V), \label{eq:soft-factor-on-eps} \\
    V_1 \cdot \left(\subsoftS{g_i} F_j \right) \cdot V_2 = & \, - \halfsoftS{g_j g_i} V_1 \cdot (F_j \cdot F_i - F_i \cdot F_j) \cdot V_2. \label{eq:soft-factor-on-F}
\end{align}
Here, $V^\mu$ represents an arbitrary Lorentz vector, and the field strength tensor $F^{\mu \nu}$ has been previously defined. 

In \cite{softYMS}, a bottom-up construction of YMS amplitudes was proposed by considering the soft behavior of amplitudes. By studying the YMS amplitudes in the soft limit, one can build an expansion formula of amplitudes from amplitudes with a chosen gluon or scalar removed. The explicit expressions of YMS amplitudes were verified from the soft behaviors of other particles and the permutation symmetry among gluons. Specifically, the expansion formula of YMS amplitudes (\ref{eq:exp-poly}) was constructed with the soft bootstrap \cite{softYMS}. Additionally, following the bottom-up construction, another form of the expansion is given by
\begin{equation} \label{eq:exp-frac}
    \yms(1, \ldots, n \|  \{g_i\}  \, | \, \bs) = \sum_{\ba} \frac{p_r \cdot F_{\ba^{\text{T}}} \cdot \leftX{\ba}}{p_r \cdot K} \yms(1, \{2, \ldots, n-1\} \shuffle \ba, n  \|  \{g_i\} \! \setminus \! \ba  \, | \, \bs),
\end{equation} 
where we sum over all ordered subsets $\ba=(\alpha_1, \ldots, \alpha_a)$ of gluons $\{g_i\}$. The arbitrary reference momentum $p_r$ is introduced here and reflects the gauge invariance of the amplitude \cite{softYMS}. The  $K$ collects the momenta of all gluons $\{g_i\}$. The  $\leftX{\ba}$ collects the momenta of scalars at the LHS of gluon $\alpha_1$. Compared to the expansion (\ref{eq:exp-poly}) with the specific selection of the fiducial gluon $g_f$, the expansion formula (\ref{eq:exp-frac}) manifestly respects both permutation symmetry among gluons and gauge invariance. However, the coefficients of the expansion formula (\ref{eq:exp-frac}) contain spurious poles without a clear physical meaning \cite{softYMS}.

\section{A recursive expansion relation from the HAB formula} \label{sec:exp-from-hopf}
In this section, we show that the HAB formulas (\ref{eq:hopf-result}), (\ref{eq:hopf-result-2s}) induce a recursive expansion relation for YMS amplitudes mixing gluons, massive scalars, and massless scalars. By this recursion, some gluons in the original amplitude are converted into massless scalars, accompanied by proper kinematic coefficients which absorb the polarization vectors. We present a simple example and then sketch the general verification for amplitudes with at least three scalars. The boundary case with only two massive scalars is further studied.


\subsection{Example}
We now demonstrate the recursive expansion relation by YMS amplitude involving three massive scalars and two gluons $\yms(1, 2, 3  \|  \{g_1, g_2\}  \, | \, \bs)$. According to the HAB formula (\ref{eq:hopf-result}), the amplitude is expressed as a sum over the two gluon orderings $(g_1, g_2)$ and $(g_2, g_1)$ as
\begin{equation}
    \yms(1, 2, 3 \|  \{g_1, g_2\}  \, | \, \bs) = \mathbf{m}(1, g_1, g_2, 2, 3  \, | \, \bs) \, \kPreN\left(1, g_1, g_2, 2, 3 \right) + (g_1 \leftrightarrow g_2).
\end{equation}
The explicit expressions of the kinematic coefficients follow from (\ref{eq:kinematic-pre-numerator}) when the ordered partitions of gluons $g_1$ and $g_2$, $\mathbf{P}_{ \!\! \br^g}^{1} \! \! : \! \! \{g_1, g_2\}$ and $\mathbf{P}_{ \!\! \br^g}^{2} \! \! : \! \! \{(g_1), (g_2)\},\{(g_2), (g_1)\}$ are taken into account:
\begin{equation} \label{eq:pre-numerator-3s2g-g1g2}
    \kPreN\left(1, g_1, g_2, 2, 3 \right) = - \frac{2 p_1 \cdot F_{12} \cdot p_2}{p_{12}^2 - m_s^2} + \frac{2 p_{1} \cdot F_1 \cdot p_2}{p_{12}^2 - m^2} \frac{2 (p_{1} + k_1) \cdot F_2 \cdot p_2}{(p_{12} + k_1)^2 - m_s^2} + \frac{2 p_{1} \cdot F_2 \cdot p_2}{p_{12}^2 - m^2} \frac{2 p_{1} \cdot F_1 \cdot (p_2 + k_2)}{(p_{12} + k_2)^2 - m_s^2}.
\end{equation}
Relabeling $g_1 \leftrightarrow g_2$, we get the other coefficient
\begin{equation} \label{eq:pre-numerator-3s2g-g2g1}
    \kPreN\left(1, g_2, g_1, 2, 3 \right) = - \frac{2 p_1 \cdot F_{21} \cdot p_2}{p_{12}^2 - m_s^2}  + \frac{2 p_{1} \cdot F_1 \cdot p_2}{p_{12}^2 - m^2} \frac{2 p_{1} \cdot F_2 \cdot (p_2 + k_1)}{(p_{12} + k_1)^2 - m_s^2} + \frac{2 p_{1} \cdot F_2 \cdot p_2}{p_{12}^2 - m^2} \frac{2 (p_{1} + k_2) \cdot F_1 \cdot p_2}{(p_{12} + k_2)^2 - m_s^2}.
\end{equation}
Note that these terms fall into two classes: (i) a single-factor term, e.g, $\frac{2 p_1 \cdot F_{12} \cdot p_2}{p_{12}^2 - m_s^2}$, in which the strength tensors of the two gluons are contracted together, and (ii) two-factor terms, e.g, $\frac{2 p_{1} \cdot F_1 \cdot p_2}{p_{12}^2 - m^2} \frac{2 (p_{1} + k_1) \cdot F_2 \cdot p_2}{(p_{12} + k_1)^2 - m_s^2}$, in which the strength tensors of the two gluons appear in different factors.

A key observation is that whenever a gluon occupies a single factor, it plays the role of a scalar from the perspective of the remaining gluon. As an illustration, consider the term $\frac{2 p_{1} \cdot F_1 \cdot p_2}{p_{12}^2 - m^2} \frac{2 (p_{1} + k_1) \cdot F_2 \cdot p_2}{(p_{12} + k_1)^2 - m_s^2}$, the first factor is occupied by gluon $g_1$, whereas the second factor contains the combined momentum $p_{1} + k_1$. Hence, from the viewpoint of $g_2$, gluon $g_1$ behaves as a scalar. Consequently, the two-factor terms can be related to YMS amplitudes in which the prior gluon has already been converted into a massless scalar. For example, the second term of the kinematic coefficient (\ref{eq:pre-numerator-3s2g-g1g2}) can be identified as a reduced kinematic coefficient with a factor
\begin{equation}
    \frac{2 p_{1} \cdot F_1 \cdot p_2}{p_{12}^2 - m^2} \frac{2 (p_{1} + k_1) \cdot F_2 \cdot p_2}{(p_{12} + k_1)^2 - m_s^2} = \frac{2 p_{1} \cdot F_1 \cdot p_2}{p_{12}^2 - m^2} \kPreN\left(1, \bg_1, g_2, 2, 3\right), 
\end{equation}
where we used  $\bg_1$ to emphasize the role of $g_1$ as a scalar. Associating this kinematic coefficient as well as the one obtained by exchanging $g_1$, $g_2$ (i.e., the second term of (\ref{eq:pre-numerator-3s2g-g2g1})) with the corresponding propagator matrix, we obtain 
\begin{equation}
    \begin{aligned}
        & \frac{2 p_{1} \cdot F_1 \cdot p_2}{p_{12}^2 - m^2}\left[\mathbf{m}(1, g_1, g_2, 2, 3  \, | \, \bs)  \frac{2 (p_{1} + k_1) \cdot F_2 \cdot p_2}{(p_{12} + k_1)^2 - m_s^2} + \mathbf{m}(1, g_2, g_1, 2, 3  \, | \, \bs) \frac{2 p_{1} \cdot F_2 \cdot (p_2 + k_1)}{(p_{12} + k_1)^2 - m_s^2}\right] \\
        = & \frac{2 p_{1} \cdot F_1 \cdot p_2}{p_{12}^2 - m^2}\Big[\mathbf{m}(1, g_1, g_2, 2, 3  \, | \, \bs)  \kPreN\left(1, \bg_1, g_2, 2, 3\right) + \mathbf{m}(1, g_2, g_1, 2, 3  \, | \, \bs) \kPreN\left(1, g_2, \bg_1, 2, 3\right)\Big] \\
        \equiv &\frac{2 p_{1} \cdot F_1 \cdot p_2}{p_{12}^2 - m^2} \yms(1, \bg_1, 2, 3 \|  \{g_2\}  \, | \, \bs).
    \end{aligned}
\end{equation}
Here, the expression inside the squarebrackets is just the YMS amplitude $\yms(1, \bg_1, 2, 3 \|  \{g_2\}  \, | \, \bs)$ with gluon $g_2$, massless scalar $\bg_1$, and massive scalars $1$, $2$, $3$.


Following a similar discussion, other terms in (\ref{eq:pre-numerator-3s2g-g1g2}) and (\ref{eq:pre-numerator-3s2g-g2g1}) also result in YMS amplitudes with fewer gluons and more massless gluons. Altogether, the original YMS amplitude turns into
\begin{equation}\label{Eq:YMSEg1}
    \begin{aligned}
        \yms(1, 2, 3 \|  \{g_1, g_2\}  \, | \, \bs) =
        & - \frac{2 p_1 \cdot F_{12} \cdot p_2}{p_{12}^2 - m_s^2} \bas(1, \bg_1, \bg_2, 2, 3  \, | \, \bs) - \frac{2 p_1 \cdot F_{21} \cdot p_2}{p_{12}^2 - m_s^2} \bas(1, \bg_2, \bg_1, 2, 3  \, | \, \bs) \\
        & + \frac{2 p_{1} \cdot F_1 \cdot p_2}{p_{12}^2 - m^2} \yms(1, \bg_1, 2, 3 \| \{g_2\} \, | \, \bs) + \frac{2 p_{1} \cdot F_2 \cdot p_2}{p_{12}^2 - m^2} \yms(1, \bg_2, 2, 3 \|  \{g_1\}  \, | \, \bs).
    \end{aligned}
\end{equation}
The BS amplitudes $\bas(1, \bg_1, \bg_2, 2, 3  \, | \, \bs)$ and $\bas(1, \bg_2, \bg_1, 2, 3  \, | \, \bs)$  are defined by the propagator matrices $\mathbf{m}(1, g_1, g_2, 2, 3  \, | \, \bs)$ and $\mathbf{m}(1, g_2, g_1, 2, 3  \, | \, \bs)$, with  massive scalars $1$, $2$, $3$ and the massless scalars $\bg_1, \bg_2$ coming from gluons $g_1$, $g_2$. The amplitudes $\yms(1, \bg_1, 2, 3 \| \{g_2\} \, | \, \bs)$ and $\yms(1, \bg_2, 2, 3 \|  \{g_1\}  \, | \, \bs)$ stand for the YMS amplitudes where $g_1$ and $g_2$ are already converted into scalars $\bg_1$ and $\bg_2$, respectively. It is not hard to see that the relation (\ref{Eq:YMSEg1}) can be arranged into the following compact form.
%
\begin{equation} \label{eq:exp-from-hopf-3s-2g}
    \yms(1, 2, 3 \|  \{g_1, g_2\}  \, | \, \bs) = \sum_{\ba} - \frac{ (-1)^{|\ba|} 2 p_1 \cdot F_{\ba} \cdot p_2}{p_{12}^2 - m_s^2} \yms(1, \overline{\ba}, 2, 3  \|  \{g_i\} \! \setminus \! \ba  \, | \, \bs).
\end{equation}
Note that YMS amplitudes with no gluon  degenerate to BS amplitudes. 

The above example makes it explicit how gluons are converted into massless scalars and how the full YMS amplitude expands into amplitudes with fewer gluons and more massless scalars. In the next subsection, we extend (\ref{eq:exp-from-hopf-3s-2g}) to a general formula with at least three massive scalars and an arbitrary number of gluons.

\subsection{General formula for YMS with at least three massive scalars}

Inspired by the example, we now provide a general recursive expansion formula for the YMS amplitudes with at least three scalars from the HAB formula (\ref{eq:hopf-result}), (\ref{eq:kinematic-pre-numerator}).

We begin with the HAB formula (\ref{eq:hopf-result}) of YMS amplitudes with $n$ massive scalars and $m$ massless gluons
\begin{equation} \label{eq:exp-from-hopf-copy1}
    \yms(1, \ldots, n-1, n \|  \{g_i\}  \, | \, \bs) = \sum_{\br} \mathbf{m}(1, \br, n-1, n  \, | \, \bs) \, \kPreN\left(1, \br, n-1, n\right),
\end{equation}
with the explicit expression  (\ref{eq:kinematic-pre-numerator}) of the kinematic coefficient
\begin{equation} \label{eq:kinematic-pre-numerator-copy1}
    \begin{aligned}
        \kPreN\left(1, \br, n-1, n\right) & = (-1)^{m} \sum_{r=1}^{m} (-1)^{r} \sum_{\ba^l \in  \mathbf{P}_{ \!\!  \br^g}^{r}} \prod_{l=1}^{r} \frac{2 \leftRhoX{\ba^l} \cdot F_{\ba^l} \cdot \rightRhoX{\ba^l} }{P_l^2 - m_s^2} \\
        & =  \sum_{r=1}^{m} \sum_{\ba^l \in  \mathbf{P}_{ \!\! \br^g}^{r}} \prod_{l=1}^{r} - \frac{ (-1)^{|\ba^l|} 2 \leftRhoX{\ba^l} \cdot F_{\ba^l} \cdot \rightRhoX{\ba^l} }{P_l^2 - m_s^2}.
    \end{aligned}
\end{equation}

As pointed out in the previous subsection, gluons in the first ordered subset $\ba^1=\bb$ are treated as scalars for the remaining gluons $\ba^2, \ldots, \ba^r$. For all terms that share the same first subset $\bb$, one can collect the remaining terms together as follows 
\begin{equation}
    \begin{aligned}
        & \left[\,\sum_{r=1}^{m} \left. \sum_{\ba^l \in \mathbf{P}_{ \!\! \br^g}^{r}} \, \prod_{l=1}^{r} - \frac{ (-1)^{|\ba^l|} 2 \leftRhoX{\ba^l} \cdot F_{\ba^l} \cdot \rightRhoX{\ba^l} }{P_l^2 - m_s^2}\,\right] \right|_{\ba^1 = \bb} \\
        = & - \frac{ (-1)^{|\bb|} 2 \leftRhoX{\bb} \cdot F_{\bb} \cdot \rightRhoX{\bb} }{P^2 - m_s^2} \sum_{t=1}^{m - \left| \bb \right|} \sum_{\bga^\ell \in  \mathbf{P}_{\!\! \br^g \setminus \bb}^{t}} \prod_{\ell =1}^{t} - \frac{ (-1)^{|\bga^\ell|} 2 \leftRhoX{\bga^\ell} \cdot F_{\bga^\ell} \cdot \rightRhoX{\bga^\ell} }{P_\ell^2 - m_s^2}
    \end{aligned}.
\end{equation}
Here, the factors associated with $\bb$ in distinct terms are identical and have been extracted from the sum. The $P$ denotes the summation of the momenta of all scalars except for $n$, whereas the $P_l$ further includes the momenta of gluons in $\bb$ and $\bga^1, \ldots, \bga^{\ell-1}$. Those same momenta of gluons in $\bb$ also enter in the $\leftRhoX{\bga^\ell}$ and $\rightRhoX{\bga^\ell}$. For the remaining gluons $\bga^\ell$, gluons in $\bb$ are treated as scalars. Consequently, the summations on the second line yield a reduced kinematic coefficient
\begin{equation}
    \kPreN(1, \br, n-1, n)|_{\overline{\bb}} = \sum_{t=1}^{m - \left| \bb \right|} \sum_{\bga^\ell \in  \mathbf{P}_{\!\! \br^g \setminus \bb}^{t}} \prod_{\ell =1}^{t} - \frac{ (-1)^{|\bga^\ell|} 2 \leftRhoX{\bga^\ell} \cdot F_{\bga^\ell} \cdot \rightRhoX{\bga^\ell} }{P_\ell^2 - m_s^2}.
\end{equation}
The bar over the subscript $\bb$ denotes that gluons in $\bb$ are effectively massless scalars in the remaining factors. We define the trivial boundary case, in which the first subset includes all gluons, by unity
\begin{equation}
    \kPreN(1, \br, n-1, n)|_{\overline{\br^g}} = 1.
\end{equation}
This corresponds to the special term of the kinematic coefficient (\ref{eq:kinematic-pre-numerator-copy1}), which involves all gluons
\begin{equation}
    - \frac{ (-1)^{|\br^g|} 2 \leftRhoX{\br^g} \cdot F_{\br^g} \cdot \rightRhoX{\br^g} }{P^2 - m_s^2}.
\end{equation}

Then, the kinematic coefficient is written as a sum over all compatible ordered subsets of gluons
\begin{equation}
    \kPreN\left(1, \br, n-1, n\right) = \sum_{\bb(\br^g)} -\frac{ (-1)^{|\bb(\br^g)|} 2 \leftRhoX{\bb(\br^g)} \cdot F_{\bb(\br^g)} \cdot \rightRhoX{\bb(\br^g)} }{P^2 - m_s^2} \kPreN(1, \br, n-1, n)|_{\overline{\bb}(\br^g)},
\end{equation}
where each term comprises a reduced kinematic coefficient multiplied by a compatible factor. Here $\bb(\br^g)$ denotes the gluon subset compatible with the overall gluon ordering $\br^g$. For example, for the overall ordering $\br = (2, g_1, g_2)$ and gluon ordering $\br^g = (g_1, g_2)$, we have $\bb(\br^g)$: $(g_1), (g_2), (g_1, g_2)$. 

As a result, the original expansion of the amplitude (\ref{eq:exp-from-hopf-copy1}) can now be viewed as 
\begin{equation} \label{eq:sum-rho-beta}
    \begin{aligned}
        & \yms(1, \ldots, n-1, n \|  \{g_i\}  \, | \, \bs)  \\
        = & \sum_{\br} \mathbf{m}(1, \br, n-1, n  \, | \, \bs)  
        \left[\,\sum_{\bb(\br^g)} - \frac{ (-1)^{|\bb(\br^g)|} 2 \leftRhoX{\bb(\br^g)} \cdot F_{\bb(\br^g)} \cdot \rightRhoX{\bb(\br^g)} }{P^2 - m_s^2} \kPreN(1, \br, n-1, n)|_{\overline{\bb}(\br^g)}\,\right].
    \end{aligned}
\end{equation}
We now rewrite the sum over $\br$ by first shuffling the scalars $\{2, \ldots, n-2\}$ and a given ordering $\br^g = \bth$ involving all gluons, and subsequently summing over all possible $\bth$. Consequently, (\ref{eq:sum-rho-beta}) becomes
\begin{equation} \label{eq:sum-theta-rho-beta}
    \begin{aligned}
        & \yms(1, \ldots, n-1, n \|  \{g_i\}  \, | \, \bs) \\
        = & \sum_{\bth} \sum_{\br(\bth)} \mathbf{m}(1, \br(\bth), n-1, n  \, | \, \bs) \, 
        \sum_{\bb(\bth)} - \frac{ (-1)^{|\bb(\bth)|} 2 \leftOrderX{\bb(\bth)}{\br(\bth)} \cdot F_{\bb(\bth)} \cdot \rightOrderX{\bb(\bth)}{\br(\bth)} }{P^2 - m_s^2} \, \kPreN(1, \br(\bth), n-1, n)|_{\overline{\bb}(\bth)}.
    \end{aligned}
\end{equation}
Here, we summed over all $\br(\bth)\in\{2, \ldots, n-2\}\shuffle\bth$ for a given $\bth$, and then summed over all possible  $\bth$ (in other words, all nonempty ordered subsets of gluons). 


An immediate observation is that one can change the ordering of the summations in (\ref{eq:sum-theta-rho-beta}) as follows. First, we pick an ordered gluon subset $\bb$ and then collect all compatible gluon orderings $\bth(\bb)$ together with the attendant shuffle of scalars $\{2, \ldots, n-2\}$ and gluons $\bth(\bb)$. As an illustration, for a subset $\bb = (g_1, g_2)$ of gluons $\{g_1, g_2, g_3\}$, the compatible gluon orderings include $(g_3, g_1, g_2)$, $(g_1, g_2, g_3)$ and $(g_1, g_3, g_2)$. That is to say, the complete collection of $\bth(\bb)$ is precisely the shuffle of ordered gluons $\bb$ and the unordered remaining gluons $\{g_i\} \! \setminus \! \bb$. Exploiting this freedom, we reorganize the series of summations (\ref{eq:sum-theta-rho-beta}) as
\begin{equation} \label{eq:sum-beta-theta-rho}
    \begin{aligned}
        & \yms(1, \ldots, n-1, n \|  \{g_i\}  \, | \, \bs) \\
        = & \sum_{\bb} \sum_{\bth(\bb)} \sum_{\br(\bth(\bb))} - \frac{ (-1)^{|\bb|} 2 \leftOrderX{\bb}{\br(\bth(\bb))} \cdot F_{\bb} \cdot \rightOrderX{\bb}{\br(\bth(\bb))} }{P^2 - m_s^2} \, \mathbf{m}(1, \br(\bth(\bb)), n-1, n  \, | \, \bs) \, \kPreN(1, \br(\bth(\bb)), n-1, n)|_{\overline{\bb}}, 
    \end{aligned}
\end{equation}
where, the three summations were taken over all $\bb$, all $\bth(\bb)\in \bb\shuffle\text{perms}(\{g_i\} \! \setminus \! \bb)$ ($\text{perms} A$ for a set $A$ denotes all permutations of elements in $A$) and all $\br(\bth(\bb))\in\{2, \ldots, n-2\}\shuffle\bth(\bb)$.

Noting that for a given $\bb$ in the first summation, the second and the third summations can be achieved by summing over all possible  $\br\in\{2, \ldots, n-2\}\shuffle \bb\shuffle\text{perms}(\{g_i\} \! \setminus \! \bb)$. According to the properties of shuffling, this summation further split into two steps (i) summing over all $\bvr(\bb) \in \{2, \ldots, n-2\} \shuffle \bb$, and then (ii) summing over $\br(\bvr(\bb)) \in \bvr(\bb) \shuffle \text{perms}(\{g_i\} \! \setminus \! \bb)$ for a given $\bvr(\bb)$ in (i). We therefore reexpress (\ref{eq:sum-beta-theta-rho}) as 

\begin{equation} \label{eq:exp-from-hopf-rho}
    \begin{aligned}
        & \yms(1, \ldots, n-1, n \|  \{g_i\}  \, | \, \bs) \\
        = & \sum_{\bb} \sum_{\bvr(\bb)} - \frac{ (-1)^{|\bb|} 2 \leftOrderX{\bb}{\bvr(\bb)} \cdot F_{\bb} \cdot \rightOrderX{\bb}{\bvr(\bb)} }{P^2 - m_s^2} 
        \sum_{\br(\bvr(\bb))} \mathbf{m}(1, \br(\bvr(\bb)), n-1, n  \, | \, \bs) \, \kPreN(1, \br(\bvr(\bb)), n-1, n)|_{\overline{\bb}}.
    \end{aligned}
\end{equation}
The last summation is nothing but the reduced YMS amplitude $\yms(1, \bvr(\bb), n-1, n \|  \{g_i\} \! \setminus \! \bb  \, | \, \bs)$, in which the gluons in $\bb$ are treated as additional massless scalars, while elements of $\{g_i\} \! \setminus \! \bb$ play as gluons. Hence, we finally arrive at the recursive expression of YMS amplitude $\yms(1, \ldots, n-1, n \|  \{g_i\}  \, | \, \bs) $:
\begin{equation}
    \begin{aligned} \label{eq:exp-from-hopf}
        & \yms(1, \ldots, n-1, n \|  \{g_i\}  \, | \, \bs)\\
        = & \sum_{\bb} \sum_{\bvr(\bb)\in\bb\shuffle\{2,\ldots,n-2\}} - \frac{ (-1)^{|\bb|} 2 \leftOrderX{\bb}{\bvr(\bb)} \cdot F_{\bb} \cdot \rightOrderX{\bb}{\bvr(\bb)} }{P^2 - m_s^2} \yms(1, \bvr(\bb), n-1, n \|  \{g_i\} \! \setminus \! \bb  \, | \, \bs).
    \end{aligned}
\end{equation}

Thus, we have constructed a new recursive expansion relation for YMS amplitudes. 

\textbf{Comments on the result (\ref{eq:exp-from-hopf})}: It is straightforward to see that the formula (\ref{eq:exp-from-hopf}) also holds when some of the scalars in the color-ordered YMS amplitude $\yms(1, \ldots, n-1, n \|  \{g_i\}  \, | \, \bs)$ are replaced by massless scalars. One therefore reconstructs the HAB formula by applying (\ref{eq:exp-from-hopf}) iteratively. 


The above discussion only concentrated on YMS amplitudes with at least three scalars (with the fixing ones $1$, $n-1$, and $n$). In the following subsection, we study the boundary case in which the YMS amplitude only involves two scalars.

\subsection{Boundary case: YMS amplitudes with two massive scalars}
Analogous to the case of YMS amplitudes with at least three scalars,  amplitudes with two scalars and $m$ gluons are also represented as a sum of  products of the kinematic coefficients and the propagator matrices. The explicit expression (\ref{eq:hopf-result-2s}) is given by
\begin{equation} 
    \yms(1, 2 \|  \{g_i\}  \, | \, \bs) = \sum_{\br} \mathbf{m}(1, \br, g_1, 2  \, | \, \bs) \kPreN \left(1, \br, g_1, 2\right),
\end{equation}
where the gluon $g_1$ plays as one of the fixed elements on the RHS. The kinematic coefficient (\ref{eq:kinematic-pre-numerator-2s}) takes the form
\begin{equation} \label{eq:kinematic-pre-numerator-2s-copy1}
    \begin{aligned}
        \kPreN(1, \br, g_1, 2) = & \, (-1)^{m} \sum_{r=1}^{m} (-1)^{r} \sum_{\ba^l \in  \mathbf{P}_{\!\! \br^g \setminus g_1}^{r}} - \frac{2 p_1 \cdot F_1 \cdot F_{\ba^1} \cdot p_1}{P_1^2 - m^2_s} \prod_{l = 2}^{m} \frac{2 \leftRhoX{\ba^l} \cdot F_{\ba^l} \cdot \rightRhoX{\ba^l}}{P_l^2 - m^2_s},\\
        = & \sum_{r=1}^{m} \sum_{\ba^l \in  \mathbf{P}_{\!\! \br^g \setminus g_1}^{r}} \frac{ (-1)^{|\ba^1|+1} 2 p_1 \cdot F_1 \cdot F_{\ba^1} \cdot p_1}{P_1^2 - m^2_s} \prod_{l = 2}^{m} - \frac{ (-1)^{|\ba^l|} 2 \leftRhoX{\ba^l} \cdot F_{\ba^l} \cdot \rightRhoX{\ba^l}}{P_l^2 - m^2_s},
    \end{aligned}
\end{equation}

Following a discussion parallel with the previous subsection, we can perform a resummation of terms over the first subset $\ba^1 = \bb$. This is permissible due to the observation that the fiducial gluon $g_1$ and the gluons in $\bb$ are effectively treated as scalars with respect to the remaining factors. The resulting factorization of the kinematic coefficient takes the form
\begin{equation}
    \kPreN(1, \br, g_1, 2) = \sum_{\bb(\br^g)} \frac{ (-1)^{|\bb(\br^g)| + 1} 2 p_1 \cdot F_{\bb(\br^g)} \cdot F_1 \cdot p_1}{P^2 - m^2_s} \kPreN(1, \br, g_1, 2)|_{\overline{\bb}(\br^g)}.
\end{equation}
Here, the reduced kinematic coefficients are constructed from the corresponding factors. The YMS amplitudes with two scalars can then be expressed as a nested double summation as follows
\begin{equation}
    \yms(1, 2 \|  \{g_i\}  \, | \, \bs) = \sum_{\br} \mathbf{m}(1, \br, g_1, 2  \, | \, \bs) \sum_{\bb(\br^g)} \frac{ (-1)^{|\bb(\br^g)| + 1} \, 2 p_1 \cdot F_{\bb(\br^g)} \cdot F_1 \cdot p_1}{P^2 - m^2_s} \kPreN(1, \br, g_1, 2)|_{\overline{\bb}(\br^g)}.
\end{equation}
Exchanging the order of the two summations yields
\begin{equation}
    \yms(1, 2 \|  \{g_i\}  \, | \, \bs) = \sum_{\bb} \frac{ (-1)^{|\bb| + 1} \, 2 p_1 \cdot F_{\bb} \cdot F_1 \cdot p_1}{P^2 - m^2_s} \sum_{\br(\bb)} \mathbf{m}(1, \br(\bb), g_1, 2  \, | \, \bs) \kPreN(1, \br(\bb), g_1, 2)|_{\overline{\bb}}.
\end{equation}
The second summation is the reduced YMS amplitude $\yms(1, \bb, g_1, 2 \|  \{g_i\} \! \setminus \! (g_1 \cup \bb)  \, | \, \bs)$ with additional massless scalars converted from $g_1$ and gluons in $\bb$. As a result, we express the amplitude $\yms(1, 2 \|  \{g_i\}  \, | \, \bs)$ with two massive scalars $1$, $2$ by amplitudes with fewer gluons and additional massless scalars,  through the following recursive expansion relation
\begin{equation} \label{eq:exp-from-hopf-2s}
    \yms(1, 2 \|  \{g_i\}  \, | \, \bs) = \sum_{\bb} \frac{ (-1)^{|\bb| + 1} 2 p_1 \cdot F_{\bb} \cdot F_1 \cdot p_1}{P^2 - m^2_s} \yms(1, \bb, g_1, 2 \|  \{g_i\} \! \setminus \! (g_1 \cup \bb)  \, | \, \bs).
\end{equation}
This formula has a form analogous to the general formula (\ref{eq:exp-from-hopf}) derived previously, but has distinct expansion coefficients.

In the next section, we reconstruct the expansion formulas (\ref{eq:exp-from-hopf}) and (\ref{eq:exp-from-hopf-2s}) by analyzing the soft behavior of the YMS amplitudes. 

\section{Understanding the recursive expansion formula by soft behavior} \label{sec:exp-from-soft}
In this section, we understand the expansion formula (\ref{eq:exp-from-hopf}) proposed in the previous section by soft behavior of YMS amplitudes, through a similar discussion with  \cite{Zhou:2022orv}. For brevity, we demonstrate this approach by a simple example, the YMS amplitude  $\yms(1, 2, 3 \| \{g_1, g_2\} \, | \, \bs)$   with three massive scalars and two gluons. General understanding just follows from a similar discussion.



We begin with the soft behavior of the YMS amplitude $\yms(1, 2, 3 \| \{g_1, g_2\} \, | \, \bs)$, where gluon $g_2$ is the soft one, i.e., $k_2 \equiv \tau \hat{k}_2$ ($\tau \to 0$) and focus on the \textit{subleading order}\footnote{One may try to begin with the leading order. However, there exists subleading contribution since the coefficient may contain $F^{\mu\nu}_{g_2}$. As demonstrated in \cite{Zhou:2022orv}, the subleading behavior reveals the full information of the expansion coefficients.} behavior:
\begin{equation} \label{eq:soft-3s-2g}
    \begin{aligned}
        \yms^{(1)_{g_2}}(1, 2, 3 \|  \{g_1, g_2\}  \, | \, \bs) & = \subsoftS{g_2} \yms(1, 2, 3 \|  \{g_1\}  \, | \, \bs \! \setminus \! g_2) \\
        & = \subsoftS{g_2} \frac{2 p_1 \cdot F_1 \cdot p_2}{p_{12}^2 - m_s^2} \yms(1, \bg_1, 2, 3  \, | \, \bs \! \setminus \! g_2) \\
        & = B_\text{A} + B_\text{N} + B_\text{D}.
    \end{aligned}
\end{equation}
The expression can be decomposed into three distinct contributions from the action of the soft factor $\subsoftS{g_2}$ on the amplitude, on the numerator, and on the denominator, denoted by $B_{\text{A}}$, $B_{\text{N}}$, and $B_{\text{D}}$, respectively. In the following, we evaluate these terms individually.

First, the term $B_{\text{A}}$ arises from the action of the soft factor of gluon $g_2$ on the reduced amplitude
\begin{equation} \label{eq:BA}
    B_\text{A} = \frac{2 p_1 \cdot F_1 \cdot p_2}{p_{12}^2 - m_s^2} \left[\subsoftS{g_2} \yms(1, \bg_1, 2, 3  \, | \, \bs \! \setminus \! g_2)\right] = \frac{2 p_1 \cdot F_1 \cdot p_2}{p_{12}^2 - m_s^2} \yms^{(1)_{g_2}}(1, \bg_1, 2, 3 \|  g_2  \, | \, \bs),
\end{equation}

Next, we consider the numerator contribution $B_{\text{N}}$. This term arises from the action of the soft factor on $p_1 \cdot F_1 \cdot p_2$. Using the properties (\ref{eq:soft-factor-on-k}), (\ref{eq:soft-factor-on-F}) of the soft factor, we have
\begin{equation}
    \begin{aligned}
        (\subsoftS{g_2} p_1) \cdot F_1 \cdot p_2 &= -\halfsoftS{1 g_2} p_1 \cdot F_2 \cdot F_1 \cdot p_2, \\
        p_1 \cdot (\subsoftS{g_2} F_1) \cdot p_2 &= - \halfsoftS{g_1 g_2} p_1 \cdot (F_1 \cdot F_2 - F_2 \cdot F_1) \cdot p_2, \\
        p_1 \cdot F_1 \cdot (\subsoftS{g_2} p_2) &= \halfsoftS{2 g_2} p_1 \cdot F_1 \cdot F_2 \cdot p_2.
    \end{aligned}
\end{equation}
Combining these terms gives the total action of the soft factor
\begin{equation}
    \subsoftS{g_2} (p_1 \cdot F_1 \cdot p_2) = \left(-\halfsoftS{g_1 g_2} + \halfsoftS{2 g_2}\right) (p_1 \cdot F_1 \cdot F_2 \cdot p_2) + \left(-\halfsoftS{1 g_2} + \halfsoftS{g_1 g_2}\right) (p_1 \cdot F_2 \cdot F_1 \cdot p_2).
\end{equation}
Utilizing the antisymmetry of $\delta_{ij}$, we reorganize the terms as
\begin{equation}
    \halfsoftS{g_1 g_2} - \halfsoftS{2 g_2} = \halfsoftS{g_1 g_2} + \halfsoftS{g_2 2} = \tau \left. \softS{g_2} \right|_{g_1 g_2 2}, 
\end{equation}
where the subscript denotes the relative order of particles as $(g_1, g_2, 2)$. This soft factor, once combined with the YMS amplitude $\yms(1, \bg_1, 2, 3  \, | \, \bs \! \setminus \! g_2)$, yields the soft behavior of the amplitude
\begin{equation}
    \yms^{(0)_{\bg_2}}(1, \bg_1, \bg_2, 2, 3  \, | \, \bs \! \setminus \! g_2) = \left. \softS{g_2} \right|_{g_1 g_2 2} \yms(1, \bg_1, 2, 3  \, | \, \bs \! \setminus \! g_2). 
\end{equation}
Here, the $g_2$ is treated as a soft scalar. This procedure works the same for $\halfsoftS{1 g_2} - \halfsoftS{g_1 g_2} = \tau \left. \softS{g_2} \right|_{1 g_2 g_1}$. Therefore, by identifying these terms as the soft behavior of reduced YMS amplitudes, the total contribution $B_\text{N}$ from the numerator becomes
\begin{equation} \label{eq:BN}
    B_\text{N} = - \tau \left[ \frac{2 p_1 \cdot F_1 \cdot F_2 \cdot p_2}{p_{12}^2 - m_s^2} \yms^{(0)_{g_2}}(1, \bg_1, \bg_2, 2, 3  \, | \, \bs) + \frac{2 p_1 \cdot F_2 \cdot F_1 \cdot p_2}{p_{12}^2 - m_s^2} \yms^{(0)_{g_2}}(1, \bg_2, \bg_1, 2, 3  \, | \, \bs) \right].
\end{equation}


Finally, we calculate the denominator contribution $B_{\text{D}}$. 
\begin{equation}
    B_{\text{D}} = - \frac{2 p_1 \cdot F_1 \cdot p_2}{(p_{12}^2 - m_s^2)^2} \left[\subsoftS{g_2} (p_{12}^2 - m_s^2)\right] \yms(1, \bg_1, 2, 3  \, | \, \bs \! \setminus \! g_2)
\end{equation}
Due to the antisymmetry of the field strength, the action of the soft factor on the momentum squared is identically zero. Consequently, the non-vanishing contribution comes from $p_1 \cdot p_2$
\begin{equation}
    \subsoftS{g_2} (p_1 \cdot p_2) = \left(- \halfsoftS{1 g_2} + \halfsoftS{2 g_2}\right)(p_1 \cdot F_2 \cdot p_2).
\end{equation}
The terms in the first bracket are rewritten as soft factors 
\begin{equation}
    \halfsoftS{1 g_2} - \halfsoftS{2 g_2} = \halfsoftS{1 g_2} + \halfsoftS{g_2 g_1} + \halfsoftS{g_1 g_2} + \halfsoftS{g_2 2} = \tau \left(\left. \softS{g_2} \right|_{1 g_2 g_1} + \left. \softS{g_2} \right|_{g_1 g_2 2}\right).
\end{equation}
Each soft factor is separately connected to the corresponding reduced YMS amplitude
\begin{align}
    \yms^{(0)_{g_2}}(1, \bg_2, \bg_1, 2, 3  \, | \, \bs) = & \left. \softS{g_2} \right|_{1 g_2 g_1} \yms(1, \bg_1, 2, 3  \, | \, \bs \! \setminus \! g_2), \\
    \yms^{(0)_{g_2}}(1, \bg_1, \bg_2, 2, 3  \, | \, \bs) = & \left. \softS{g_2} \right|_{g_1 g_2 2} \yms(1, \bg_1, 2, 3  \, | \, \bs \! \setminus \! g_2).
\end{align}
Thus, $B_\text{D}$ becomes
\begin{equation}
    B_\text{D} = \tau \frac{2 p_1 \cdot F_2 \cdot p_2}{p_{12}^2 - m_s^2} \frac{2 p_1 \cdot F_1 \cdot p_2}{p_{12}^2 - m_s^2}  \left(\yms^{(0)_{g_2}}(1, \bg_2, \bg_1, 2, 3  \, | \, \bs) + \yms^{(0)_{g_2}}(1, \bg_1, \bg_2, 2, 3  \, | \, \bs)\right).
\end{equation}
Part of the obtained expression of $B_\text{D}$ reproduces the soft behavior of a YMS amplitude with the expansion
\begin{equation}
    \begin{aligned}
        \yms(1, \bg_2, 2, 3 \| g_1 \, | \, \bs)= & \, \frac{2 (p_1+k_2) \cdot F_1 \cdot p_2}{(p_{12}+k_2)^2 - m_s^2} \yms(1, \bg_2, \bg_1, 2, 3  \, | \, \bs) \\
        & \, + \frac{2 p_1 \cdot F_1 \cdot (k_2 + p_2)}{(p_{12}+k_2)^2 - m_s^2} \yms(1, \bg_1, \bg_2, 2, 3  \, | \, \bs).
    \end{aligned}
\end{equation}
Specifically, the leading order $\mathcal{O}(\tau^{-1})$ of the soft limit takes the form of
\begin{equation}
    \yms^{(0)_{g_2}}(1, \bg_2, 2, 3 \| g_1 \, | \, \bs) = \frac{2 p_1 \cdot F_1 \cdot p_2}{p_{12}^2 - m_s^2}  \left(\yms^{(0)_{g_2}}(1, \bg_2, \bg_1, 2, 3  \, | \, \bs) + \yms^{(0)_{g_2}}(1, \bg_1, \bg_2, 2, 3  \, | \, \bs)\right),
\end{equation}
which arises from the $\mathcal{O}(\tau^{0})$ term of coefficient and the $\mathcal{O}(\tau^{-1})$ terms of reduced YMS amplitudes. Therefore, $B_\text{D}$ can be represented compactly as
\begin{equation} \label{eq:BD}
    B_\text{D} = \tau \frac{2 p_1 \cdot F_2 \cdot p_2}{p_{12}^2 - m_s^2} \yms^{(0)_{g_2}}(1, \bg_2, 2, 3 \| g_1 \, | \, \bs).
\end{equation}

Collecting the results for $B_{\text{A}}$ (\ref{eq:BA}), $B_{\text{N}}$ (\ref{eq:BN}), and $B_{\text{D}}$ (\ref{eq:BD}), the total soft behavior of the amplitude (\ref{eq:soft-3s-2g}) is given by
\begin{equation}
    \begin{aligned}
        & \yms^{(1)_{g_2}}(1, 2, 3 \|  \{g_1, g_2\}  \, | \, \bs) \\
        = & \, \frac{2 p_1 \cdot F_1 \cdot p_2}{p_{12}^2 - m_s^2} \yms^{(1)_{g_2}}(1, \bg_1, 2, 3 \|  g_2  \, | \, \bs) + \tau \frac{2 p_1 \cdot F_2 \cdot p_2}{p_{12}^2 - m_s^2} \yms^{(0)_{g_2}}(1, \bg_2, 2, 3 \| g_1 \, | \, \bs) \\
        & \, - \tau \left[ \frac{2 p_1 \cdot F_1 \cdot F_2 \cdot p_2}{p_{12}^2 - m_s^2} \yms^{(0)_{g_2}}(1, \bg_1, \bg_2, 2, 3  \, | \, \bs) + \frac{2 p_1 \cdot F_2 \cdot F_1 \cdot p_2}{p_{12}^2 - m_s^2} \yms^{(0)_{g_2}}(1, \bg_2, \bg_1, 2, 3  \, | \, \bs) \right].
    \end{aligned}
\end{equation}
This result exactly matches the soft limit of the following proposed expansion
\begin{equation}
    \begin{aligned}
        & \yms(1, 2, 3 \|  \{g_1, g_2\}  \, | \, \bs) \\
        = & \, \frac{2 p_1 \cdot F_1 \cdot p_2}{p_{12}^2 - m_s^2} \yms(1, \bg_1, 2, 3 \|  g_2  \, | \, \bs) + \frac{2 p_1 \cdot F_2 \cdot p_2}{p_{12}^2 - m_s^2} \yms(1, \bg_2, 2, 3 \| g_1 \, | \, \bs) \\
        & \, - \frac{2 p_1 \cdot F_1 \cdot F_2 \cdot p_2}{p_{12}^2 - m_s^2} \yms(1, \bg_1, \bg_2, 2, 3  \, | \, \bs) - \frac{2 p_1 \cdot F_2 \cdot F_1 \cdot p_2}{p_{12}^2 - m_s^2} \yms(1, \bg_2, \bg_1, 2, 3  \, | \, \bs).
    \end{aligned}
\end{equation}
This expression is identical to the expansion formula (\ref{eq:exp-from-hopf-3s-2g}) 
\begin{equation} 
    \yms(1, 2, 3 \|  \{g_1, g_2\}  \, | \, \bs) = \sum_{\ba} - \frac{ (-1)^{|\ba|} 2 p_1 \cdot F_{\ba} \cdot p_2}{p_{12}^2 - m_s^2} \yms(1, \overline{\ba}, 2, 3  \|  \{g_i\} \! \setminus \! \ba  \, | \, \bs).
\end{equation}
which is constructed from the Hopf-algebra-based (HAB) formula. The validity of this expansion can be further verified by considering the soft limit of the gluon $g_1$. 

The above derivation can be straightforwardly extended to the general case involving at least three scalars and an arbitrary number of gluons. Starting from the base case with one gluon, one can iteratively add a single gluon at each inductive step, systematically building the expansion of the most general amplitude by induction.

\section{The relationship between (\ref{eq:exp-poly}) and the HAB formulas} \label{sec:massless-expansion}
In this section, we establish the relationship between the recursive expansion (\ref{eq:exp-poly}), and the formulas (\ref{eq:exp-from-hopf}), (\ref{eq:exp-from-hopf-2s}) derived from the HAB formula in the massless limit. To systematically study this relationship, we proceed from the simplest cases to  more complicated ones, progressively. Particularly, we begin with the boundary case of two-scalar amplitudes. Next, we consider the base case of single-gluon amplitudes with an arbitrary number of scalars. Finally, we explicitly compute the five-point amplitude with two gluons, demonstrating that the recursive expansion (\ref{eq:exp-poly}) combined with the BCJ relations indeed reproduces the structure expected from the HAB formula. In each case, we derive a new expression from (\ref{eq:exp-poly}) and show that it matches the massless limit of the HAB formulas.

\subsection{The boundary case: amplitudes with two scalars}
The YMS amplitudes involving \textbf{massless} scalars admit an expansion in terms of amplitudes with more scalars and fewer gluons, multiplied by kinematic coefficients. We show the relation between the two expansion formulas (\ref{eq:exp-poly}) and (\ref{eq:exp-frac}) in the appendix \ref{sec:relation}. We now restrict the expression (\ref{eq:exp-frac-with-fiducial-gluon}), derived from (\ref{eq:exp-poly}), to YMS amplitudes containing exactly two scalars
\begin{equation} \label{eq:exp-poly-2s}
    \begin{aligned}
        \yms(1, 2 \|  \{g_i\}  \, | \, \bs) = & \,  \frac{p_r \cdot F_1 \cdot p_1}{p_r \cdot k_1} \yms(1, g_1, 2  \|  \{g_i\} \! \setminus \! g_1  \, | \, \bs) \\
        & + \sum_{\ba}  \frac{p_r \cdot F_1 \cdot F_{\ba^{\text{T}}} \cdot p_1}{p_r \cdot k_1} \yms(1, \ba, g_1, 2  \|  \{g_i\} \! \setminus \! (g_1 \cup \ba)  \, | \, \bs).
    \end{aligned}
\end{equation}
Due to the gauge invariance, the reference momentum $p_r$ can be chosen arbitrarily. We choose $p_r = 2 p_1$, for which the coefficient of the first term in (\ref{eq:exp-poly-2s}) vanishes. The factor of $2$ is introduced for convenience, and it allows the denominator to be rewritten as a scalar propagator $2 p_1 \cdot k_1 = (p_1 + k_1)^2$. As a result, we obtain
\begin{equation} \label{eq:exp-frac-fix3-2s}
    \yms(1, 2 \|  \{g_i\}  \, | \, \bs) = \sum_{\ba} \frac{2 p_1 \cdot F_1 \cdot F_{\ba^{\text{T}}} \cdot p_1}{(p_1 + k_1)^2} \yms(1, \ba, g_1, 2  \|  \{g_i\} \! \setminus \! (g_1 \cup \ba)  \, | \, \bs).
\end{equation}
This precisely reproduces the massless limit (i.e. $m_s^2->0$) of the recursive expansion formula (\ref{eq:exp-from-hopf-2s}) derived from the Hopf-algebra-based result, up to a reversal in the ordering of the field strength tensors 
\begin{equation}
    p_1 \cdot F_1 \cdot F_{\ba^{\text{T}}} \cdot p_1 = (-1)^{|\ba|+1} p_1 \cdot F_{\ba} \cdot F_1 \cdot p_1,
\end{equation}
due to the antisymmetry property of the field strength tensors $F^{\mu \nu} = - F^{\nu \mu}$.

Having verified the equivalence for the boundary case, we now turn to amplitudes involving gluons. The simplest case involves a single gluon and multiple scalars, which serves as the base case for the recursive construction.

\subsection{The base case: amplitudes with one gluon}
In this subsection, we study YMS amplitudes involving at least three massless scalars and a single gluon.

Consider the general expansion formula (\ref{eq:exp-frac}) restricted to amplitudes with only one gluon,
\begin{equation} \label{eq:exp-frac-with-fiducial-gluon-1g}
    \yms(1, 2, \ldots, n-1, n \|  g_1  \, | \, \bs) = \frac{p_r \cdot F_1 \cdot \leftX{}}{p_r \cdot k_1} \yms(1, \{2, \ldots, n-1\} \shuffle g_1, n  \, | \, \bs) 
\end{equation}

Exploiting gauge invariance, we choose the reference momentum to be $p_r = 2p_n$. 
We define $\rightX{}$ as the sum of the momenta of scalars to the right of the gluon $g_1$, {\bf excluding $n$}. Then the numerators of the coefficients in the expansion (\ref{eq:exp-frac-with-fiducial-gluon-1g}) can be rewritten in a more symmetric form
\begin{equation} 
    2 p_n \cdot F_1 \cdot \leftX{} = -2(k_1 + \leftX{} + \rightX{}) \cdot F_1 \cdot \leftX{} = -2\rightX{} \cdot F_1 \cdot \leftX{} = 2\leftX{} \cdot F_1 \cdot \rightX{}.
\end{equation}
Here, the momentum conservation $\leftX{} + k_1 + \rightX{} + p_n = 0$ and the antisymmetry property of the field strength tensor $F_1^{\mu \nu}$ are applied. Specifically, the coefficient associated with the YMS amplitude $\yms(1, \{2, \ldots, n-1\}, g_1, n  \, | \, \bs)$ vanishes
\begin{equation}
    2 p_n \cdot F_1 \cdot (p_1 + \ldots + p_{n-1}) = 2 p_n \cdot F_1 \cdot (- k_1 - p_n) = 0,
\end{equation}
or more directly, as a consequence of $\rightX{} = 0$. Hence, only permutations of the form $(1, \{2, \ldots, n-2\} \shuffle g_1, n-1, n)$ contribute to the expansion (\ref{eq:exp-frac-with-fiducial-gluon-1g}) of amplitudes, which implies that three scalars $1$, $n-1$ and $n$ are held fixed. 

In parallel, the denominator of the coefficients in (\ref{eq:exp-frac-with-fiducial-gluon-1g}) can be rewritten as
\begin{equation}
    2 p_n \cdot k_1 = (p_n + k_1)^2 = P^2,
\end{equation}
where $P \equiv p_1 + \ldots + p_{n-1}$ denotes the total momentum of all scalars except for $n$, namely $P = \leftX{} + \rightX{}$.

Collecting the above results, we obtain
\begin{equation} \label{eq:exp-frac-fix3-1g}
    \yms(1, 2, \ldots, n-1, n \|  g_1  \, | \, \bs) = \frac{2\leftX{} \cdot F_1 \cdot \rightX{}}{P^2} \yms(1, \{2, \ldots, n-2\} \shuffle g_1, n-1, n  \, | \, \bs) 
\end{equation}
This result is consistent with the recursive expansion formula (\ref{eq:exp-from-hopf}) derived from the Hopf-algebra-based result. This serves as the base case for our construction of YMS amplitudes by induction.

\subsection{Amplitudes with two gluons}
We take the five-point amplitude $\yms(1, 2, 3 \|  \{g_1,g_2\}  \, | \, \bs)$ with two gluons as our illustrative example which can be straightforwardly  generalized to  amplitude with two gluons and more than three scalars. Amplitudes with more than two gluons can be calculated in a similar way, but will not be presented in th current work.

When applying (\ref{eq:exp-poly}) iteratively, we express $\yms(1, 2, 3 \|  \{g_1,g_2\}  \, | \, \bs)$ by a combination of BS amplitudes as follows
\begin{equation}
\yms(1, 2, 3 \|  \{g_1,g_2\}  \, | \,\bs)=T_1+T_2,
\end{equation}
where $T_1$ and $T_2$ refer to the terms involving $(\epsilon_1\cdot\epsilon_2)^0$ and $(\epsilon_1\cdot\epsilon_2)^1$, respectively. More explicitly, they are given as 
\begin{equation}
\begin{aligned}
T_1=&\Bigl[(\epsilon_1\cdot p_1)[\epsilon_2\cdot (p_1+k_1)]\bas(1, g_1, g_2, 2, 3 \, | \,\bs)+(\epsilon_1\cdot p_1)[\epsilon_2\cdot (p_{12}+k_1)]\bas(1, g_1,2, g_2, 3 \, | \,\bs)\\
&+(\epsilon_1\cdot p_{12})[\epsilon_2\cdot (p_{12}+k_1)]\bas(1,2, g_1, g_2, 3 \, | \,\bs)\Bigr]+(g_1\leftrightarrow g_2 | \bs),\\
T_2=&-\frac{1}{2}(\epsilon_1\cdot \epsilon_2)\Bigl[(k_1\cdot p_1)\bas(1, g_1, g_2, 2, 3 \, | \,\bs)+(k_1\cdot p_1)\bas(1, g_1, 2, g_2, 3 \, | \,\bs)\\
&+(k_1\cdot p_{12})\bas(1,  2,g_1, g_2, 3 \, | \,\bs)\Bigr]+(g_1\leftrightarrow g_2|\bs).
\end{aligned}
\end{equation}
In the above expression, $(g_1\leftrightarrow g_2|\bs)$ means the term obtained by exchanging the roles of  $g_1$ and $g_2$ in the coefficients and in the left permutation, with keeping the right permutation as $\bs$.

\textit{The $T_1$ sector}~~Now we focus on $T_1$. Using BCJ relations (\ref{eq:BCJrelation1g}, \ref{eq:BCJrelation2g}) \cite{Chen:2011jxa,Bjerrum-Bohr:2009ulz} in appendix \ref{sec:BCJrelation}\footnote{An alternative approach to expanding YMS amplitudes in terms of BS ones in BCJ basis can be found in \cite{Feng:2020jck}}, one can transform all BS amplitudes  in $T_1$ into a combination of $\bas(1, g_1, g_2, 2, 3 \, | \,\bs)$ and $\bas(1, g_2, g_1, 2, 3 \, | \,\bs)$. Then $T_1$ turns into 
\begin{equation}
T_1=\left(C_1+C_2+C_3+C_4+C_5\right)\bas(1, g_1, g_2, 2, 3 \, | \,\bs)+(g_1\leftrightarrow g_2|\bs),
\end{equation}
where the coefficients are displayed as
\begin{equation}
\begin{aligned}
C_1=&(\epsilon_1\cdot p_1)\,[\epsilon_2\cdot (p_1+k_1)]+(\epsilon_1\cdot p_1)\,[\epsilon_2\cdot(p_{12}+k_1)]\,\frac{2k_2\cdot(p_1+k_1)}{(p_{12}+k_1)^2}\\
C_2=&(\epsilon_1\cdot p_{12})\,[\epsilon_2\cdot(p_{12}+k_1)]\,\frac{2k_1\cdot p_1}{p_{12}^2}+(\epsilon_1\cdot p_{12})\,[\epsilon_2\cdot(p_{12}+k_1)]\,\frac{2k_1\cdot p_1}{p_{12}^2}\frac{2k_2\cdot (p_1+k_1)}{(p_{12}+k_1)^2}\\
C_3=&(\epsilon_1\cdot p_{12})\,[\epsilon_2\cdot(p_{12}+k_1)]\,\frac{2k_2\cdot (p_1+k_1)}{p_{12}^2}+(\epsilon_1\cdot p_{12})\,[\epsilon_2\cdot(p_{12}+k_1)]\,\frac{-2k_2\cdot (p_1+k_1)}{p_{12}^2}\\
C_4=&(\epsilon_2\cdot p_1)\,[\epsilon_1\cdot(p_{12}+k_2)]\,\frac{2k_1\cdot p_1}{(p_{12}+k_2)^2}+(\epsilon_2\cdot p_{12})\,[\epsilon_1\cdot(p_{12}+k_2)]\,{\frac{2k_2 \cdot p_1}{p_{12}^2}}\,\frac{2k_1\cdot p_1}{(p_{12}+k_2)^2}\\
C_5=&(\epsilon_2\cdot p_{12})\,[\epsilon_1\cdot(p_{12}+k_2)]\,\frac{-2k_1\cdot p_1}{p_{12}^2}.
\end{aligned}
\end{equation}
We now calculate $C_1$ as a typical example
\begin{equation}
\begin{aligned}
C_1=&(\epsilon_1\cdot p_1)\,\frac{-2(p_{12}+k_1)\cdot k_2\,\epsilon_2\cdot (p_1+k_1)+2(p_{12}+k_1)\cdot\epsilon_2 \, k_2\cdot(p_1+k_1)}{(p_{12}+k_1)^2}\\
=&(\epsilon_1\cdot p_1)\,\frac{-2p_2\cdot F_2\cdot (p_1+k_1)}{(p_{12}+k_1)^2},
\end{aligned}
\end{equation}
where we have used the fact $(p_{12}+k_1)^2=(k_2+p_3)^2=2k_2\cdot p_3=-2k_2\cdot (p_{12}+k_1)$ on the first line, due to momentum conservation and the massless condition. The antisymmetry of the field strength tensor has been considered on the second line. Using similar techniques, we simplify $C_2$, $C_3$ and $C_4$ as 
\begin{equation}
\begin{aligned}
C_2=&\,(-1)\,(\epsilon_1\cdot p_{12})\,[\epsilon_2\cdot(p_{12}+k_1)]\,\frac{2k_2\cdot p_2}{(p_{12}+k_1)^2}\,\frac{2k_1\cdot p_1}{p_{12}^2}, \quad C_3=0,\\
C_4=&\,(-1)\,\frac{2 p_2\cdot F_2\cdot p_1}{p_{12}^2}\,\frac{2 \epsilon_1\cdot (p_{12}+k_2)\, k_1\cdot p_1}{(p_{12}+k_2)^2}+\frac{2 \epsilon_2\cdot p_1\,\epsilon_1\cdot(p_{12}+k_2)\, k_1\cdot p_1}{p_{12}^2}.
\end{aligned}
\end{equation}
Further multiplying $1=\frac{p_{12}^2}{p_{12}^2}=-\frac{2k_1\cdot p_{12}}{p_{12}^2}+\frac{2k_2\cdot p_3}{p_{12}^2}$ to $C_1$ and then take the sum of $C_1, \ldots, C_5$, we finally get 
\begin{equation}
\begin{aligned}
C_1+C_2+C_3+C_4+C_5=&\frac{2p_1\cdot F_2\cdot p_2}{p_{12}^2} \, \frac{2p_1\cdot F_1\cdot (p_2+k_2)}{(p_{12}+k_2)^2}+ \frac{2p_1\cdot F_1\cdot p_2}{p_{12}^2} \, \frac{2(p_1+k_1)\cdot F_2\cdot p_2}{(p_{12}+k_1)^2} \\
&-\frac{2(p_1\cdot F_1\cdot F_2\cdot p_2)|_{(\epsilon_1\cdot\epsilon_2)^0}}{p_{12}^2}.
\end{aligned}
\end{equation}
Here $(p_1\cdot F_1\cdot F_2\cdot p_2)|_{(\epsilon_1\cdot\epsilon_2)^0}$ denotes the terms involving $(\epsilon_1\cdot\epsilon_2)^0$ as
\begin{equation} 
(p_1\cdot F_1\cdot F_2\cdot p_2)|_{(\epsilon_1\cdot\epsilon_2)^0}= p_1\cdot k_1\,\epsilon_1\cdot k_2\,\epsilon_2\cdot p_2- p_1\cdot \epsilon_1\,k_1\cdot F_2 \cdot p_2= p_1\cdot F_1\cdot k_2\,\epsilon_2\cdot p_2 + p_1\cdot\epsilon_1\,k_1\cdot\epsilon_2\,k_2\cdot p_2.
\end{equation}
With the above coefficients and the expression for amplitude with one gluon, we reexpress $T_1$ in a convenient form 
\begin{equation}
\begin{aligned}
T_1=&\frac{2p_1\cdot F_1\cdot p_2}{p_{12}^2}\,\yms(1, g_1, 2, 3 \|  \{g_2\}  \, | \,\bs)+\frac{2p_1\cdot F_2\cdot p_2}{p_{12}^2}\,\yms(1, g_2, 2, 3 \|  \{g_1\}  \, | \,\bs)\label{eq:T1recursion}\\
&-\frac{2(p_1\cdot F_1\cdot F_2\cdot p_2)|_{(\epsilon_1\cdot\epsilon_2)^0}}{p_{12}^2}\bas(1,g_1, g_2,2, 3 \, | \,\bs)-\frac{2(p_1\cdot F_2\cdot F_1\cdot p_2)|_{(\epsilon_1\cdot\epsilon_2)^0}}{p_{12}^2}\bas(1,g_2, g_1, 2, 3 \, | \,\bs).
\end{aligned}
\end{equation}

\textit{The $T_2$ sector}~~When following an analogous discussion with the $T_1$ sector, by expressing the BS amplitudes in $T_2$ in to minimal basis $\bas(1,g_1, g_2, 2, 3 \, | \,\bs)$ and  $\bas(1,g_2, g_1, 2, 3 \, | \,\bs)$, we arrive at
\begin{equation}
T_2=-\frac{2(p_1\cdot F_1\cdot F_2\cdot p_2)|_{(\epsilon_1\cdot\epsilon_2)^1}}{p_{12}^2}\bas(1,g_1, g_2,2, 3 \, | \,\bs)-\frac{2(p_1\cdot F_2\cdot F_1\cdot p_2)|_{(\epsilon_1\cdot\epsilon_2)^1}}{p_{12}^2}\bas(1,g_2, g_1, 2, 3 \, | \,\bs),\label{eq:T2}
\end{equation}
in which, $(p_1\cdot F_1\cdot F_2\cdot p_2)|_{(\epsilon_1\cdot\epsilon_2)^1}$ stands for the $(\epsilon_1\cdot\epsilon_2)^1$ sector of $p_1\cdot F_1\cdot F_2\cdot p_2$:
\begin{equation}
(p_1\cdot F_1\cdot F_2\cdot p_2)|_{(\epsilon_1\cdot\epsilon_2)^1}=-p_1\cdot k_1\,\epsilon_1\cdot\epsilon_2\,k_2\cdot p_2.
\end{equation}
This result can also be straightforwardly obtained by considering {\it gauge invariance condition}:
\begin{equation}
\begin{aligned}
0=\yms(1, 2, 3 \|  \{g_1,g_2\}  \, | \,\bs)|_{\epsilon_2\to k_2}=T_1|_{\epsilon_2\to k_2}+T_2|_{\epsilon_2\to k_2}\label{eq:GaugeCondition}
\end{aligned}
\end{equation}
Noting that in the expression (\ref{eq:T1recursion}) of $T_1$, both amplitude $\yms(1, g_1, 2, 3 \|  \{g_2\}  \, | \,\bs)$ and coefficients containing the field strength tensor $F_2^{\mu\nu}$ are gauge invariant objects, the only surviving terms in $T_1|_{\epsilon_2\to k_2}$ are given by
\begin{equation}
T_1|_{\epsilon_2 \to k_2}=-\frac{2 p_1\cdot k_1\,\epsilon_1\cdot k_2\,k_2\cdot p_2}{p_{12}^2}\,\bas(1,g_1, g_2,2, 3 \, | \,\bs)-\frac{2 p_1\cdot k_2\,k_2\cdot\epsilon_1\,k_1\cdot p_2}{p_{12}^2} \,\bas(1,g_2, g_1,2, 3 \, | \,\bs).
\end{equation}
Substituting the above expression into (\ref{eq:GaugeCondition}), one can solve out the coefficient of $\epsilon_1\cdot {k_2}$ in $T_2|_{\epsilon_2 \to k_2}$ and verify (\ref{eq:T2}).

Summing (\ref{eq:T1recursion}) and (\ref{eq:T2}) together, we finally get 
\begin{equation}
\begin{aligned}
\yms(1, 2, 3 \| \{g_1,g_2\}  \, | \,\bs)=&\frac{2p_1\cdot F_1\cdot p_2}{p_{12}^2}\,\yms(1, g_1, 2, 3 \|  \{g_2\}  \, | \,\bs)+\frac{2p_1\cdot F_2\cdot p_2}{p_{12}^2}\,\yms(1, g_2, 2, 3 \|  \{g_1\}  \, | \,\bs)\label{eq:T1recursion}\\
&-\frac{2p_1\cdot F_1\cdot F_2\cdot p_2}{p_{12}^2}\bas(1,g_1, g_2,2, 3 \, | \,\bs)-\frac{2p_1\cdot F_2\cdot F_1\cdot p_2}{p_{12}^2}\bas(1,g_2, g_1, 2, 3 \, | \,\bs),
\end{aligned}
\end{equation}
which is just the expected massless limit of the recursion formula (\ref{eq:exp-from-hopf}) based on HAB formula.

These explicit computations—from the boundary two-scalar case to the two-gluon five-point example—collectively demonstrate that the recursive expansion (\ref{eq:exp-poly}) is consistent with the massless limit of the HAB formula. While we have not presented a general proof for an arbitrary numberof gluons, the pattern observed in these representative cases strongly suggests that the equivalence holds generally. The key mechanism appears to be that the BCJ relations systematically reorganize the coefficients involving the polarization vectors into the form of the HAB formula.

\section{Conclusions} \label{sec:conclusion}

In this work, we established the relationship between the Hopf-algebra based (HAB) formula for massive-scalar YMS amplitudes and the recursive expansion formula for massless scalar YMS amplitudes. We proposed a convenient recursive formula for massive-scalar YMS amplitudes, and confirm this formula by soft behavior approach. This recursive formula iteratively results in the HAB formula. On the other hand, the massless limit becomes a formula for massless scalar YMS amplitudes that is further derived from the recursive formula in \cite{expEYM}. We hope this work could provide a new insight into the study of matter coupling to gravitation.



\section*{Acknowledgments}
We would like to thank Kang Zhou, Gang Chen, Chih-Hao Fu, Yihong Wang and Chongsi Xie for helpful discussions. This work is supported by NSFC under Grant No. 11875206.

\appendix

\section{Useful BCJ relations} \label{sec:BCJrelation}
The fundamention BCJ relation for BS amplitude is given as 
\begin{equation} 
\bas(1,\ldots,n-1,g_1,n\, | \,\bs)=\sum_{\br \in \{g_1\}\shuffle\{2,\ldots,n-2\}}\frac{2 k_1\cdot \leftRhoX{g_1}}{(k_{1}+ p_n)^2}\,\bas(1,\{g_1\}\shuffle\{2,\ldots,n-2\},n-1,n \, | \,\bs).\label{eq:BCJrelation1g}
\end{equation}
A more complicated BCJ relation used in this paper is 
\begin{equation} 
\begin{aligned}
&\bas(1,\ldots,n-1,g_1,g_2,n\, | \,\bs)\\
=&\sum_{\br\in \{g_1 ,g_2\}\shuffle\{2,\ldots,n-2\}}\frac{2k_1\cdot \leftRhoX{g_1}+2k_2\cdot \leftRhoX{g_2}}{(k_{1}+k_2+p_n)^2}\,\bas(1,\{g_1,g_2\} \shuffle\{2,\ldots,n-2\},n-1,n\, | \,\bs)\\
&+\sum_{\br\in \{g_1\}\shuffle\{2,\ldots,n-2\}}\frac{2k_1\cdot \leftRhoX{g_1}+2k_2\cdot (p_{1 \ldots n-1}+k_1)}{(k_{1}+k_2+p_n)^2}\,\bas(1,\{g_1\} \shuffle \{2,\ldots,n-2\},n-1, g_2, n\, | \,\bs).\label{eq:BCJrelation2g}
\end{aligned}
\end{equation}
Apparently, by combining (\ref{eq:BCJrelation1g}) and (\ref{eq:BCJrelation2g}), one can express all amplitudes $\bas(1,\ldots,n-1,g_1,g_2,n\, | \,\bs)$ and $\bas(1,\ldots,g_1,\ldots,n-1,g_2,n\, | \,\bs)$ in terms of $\bas(1,\{g_1\}\shuffle\{g_2\}\shuffle\{2,\ldots,n-2\},n-1,n\, | \,\bs)$.

\section{{Relation between the two expansion formulas (\ref{eq:exp-poly}) and (\ref{eq:exp-frac})}}
\label{sec:relation}
We begin with the formula (\ref{eq:exp-poly}) here 
\begin{equation} \label{eq:exp-poly-copy1}
    \begin{aligned}
        \yms(1, \ldots, n \|  \{g_i\}  \, | \, \bs) = & \, \left(\epsilon_f \cdot \leftX{f}\right) \yms(1, \{2, \ldots, n-1\} \shuffle g_f, n  \|  \{g_i\} \! \setminus \! g_f  \, | \, \bs) \\
        & + \sum_{\ba}\left(\epsilon_f \cdot F_{\ba^{\text{T}}} \cdot \leftX{\ba}\right) \yms(1, \{2, \ldots, n-1\} \shuffle \{\ba, g_f\}, n  \|  \{g_i\} \! \setminus \! (g_f \cup \ba )  \, | \, \bs),
    \end{aligned}
\end{equation}
which contains only polynomial coefficients. The gauge invariance condition $ \yms|_{\epsilon_f\to k_f}=0$ associating with the fiducial gluon $g_f$ implies the following identity
\begin{equation} \label{eq:exp-poly-replace-eps-to-k}
    \begin{aligned}
        0 = & \, \left(k_f \cdot \leftX{f}\right) \yms(1, \{2, \ldots, n-1\} \shuffle g_f, n  \|  \{g_i\} \! \setminus \! g_f  \, | \, \bs) \\
        & + \sum_{\ba}\left(k_f \cdot F_{\ba^{\text{T}}} \cdot \leftX{\ba}\right) \yms(1, \{2, \ldots, n-1\} \shuffle \{\ba, g_f\}, n  \|  \{g_i\} \! \setminus \! (g_f \cup \ba )  \, | \, \bs).
    \end{aligned}
\end{equation}
Using (\ref{eq:exp-poly-copy1}) and (\ref{eq:exp-poly-replace-eps-to-k}), we can absorb the polarization vector $\epsilon_f$ into the field strength tensor $F_f^{\mu\nu}$. To this end, we introduce  an arbitrarily chosen reference momentum $p_r^\mu$ and note the following relation
\begin{equation} \label{eq:pr-dot-F}
    \frac{(p_{r})_{\mu} F_f^{\mu \nu}}{p_r \cdot k_f} = \epsilon_f^{\nu} - \frac{p_r \cdot \epsilon_f}{p_r \cdot k_f} k_f^{\nu}.
\end{equation}
The recursive expansion formula (\ref{eq:exp-poly-copy1}), the identity (\ref{eq:exp-poly-replace-eps-to-k}) and the relation (\ref{eq:pr-dot-F}) together induce the formula
\begin{equation} \label{eq:exp-frac-with-fiducial-gluon}
    \begin{aligned}
        \yms(1, \ldots, n \|  \{g_i\}  \, | \, \bs) = & \, \frac{p_r \cdot F_f \cdot \leftX{f}}{p_r \cdot k_f} \yms(1, \{2, \ldots, n-1\} \shuffle g_f, n  \|  \{g_i\} \! \setminus \! g_f  \, | \, \bs) \\
        & + \sum_{\ba}\frac{p_r \cdot F_f \cdot F_{\ba^{\text{T}}} \cdot \leftX{\ba}}{p_r \cdot k_f} \yms(1, \{2, \ldots, n-1\} \shuffle \{\ba, g_f\}, n  \|  \{g_i\} \! \setminus \! (g_f \cup \ba )  \, | \, \bs),
    \end{aligned}
\end{equation}
where the coefficients involve nontrivial denominators now.
This result is highly analogous to the expansion formula (\ref{eq:exp-frac}) proposed in \cite{softYMS}. There are, however, two important differences: \textit{First}, the former expression (\ref{eq:exp-frac-with-fiducial-gluon}) explicitly involves a fiducial gluon $g_f$, thereby breaking permutation symmetry among gluons, which is kept in the latter expression  (\ref{eq:exp-frac}). Averaging (\ref{eq:exp-frac-with-fiducial-gluon}) over all gluons yields (\ref{eq:exp-frac}). \textit{Second}, the $k_f$ in (\ref{eq:exp-frac-with-fiducial-gluon}) refers to the momentum of the fiducial gluon $g_f$, whereas $K$ in  (\ref{eq:exp-frac}) collects the momenta of all gluons. 

\bibliographystyle{JHEP}
\bibliography{ref.bib}

\end{document}